\documentclass[twocolumn]{aastex62}

\hypersetup{linkcolor=red,citecolor=blue,filecolor=cyan,urlcolor=magenta}

\usepackage{braket}
\usepackage{amsmath}
\usepackage{bm}

\newcommand{\lya}{Ly$\alpha\,$}

\newcommand{\taueff}{$\tau_{\rm eff}\,$}
\newcommand{\cmpch}{cMpc$\,h^{-1}\,$}
\newcommand{\kms}{km\,s$^{-1}$}

\shorttitle{Opacity of the Intergalactic Medium at $z\sim 6$}
\shortauthors{A.-C. Eilers et al.}

\begin{document}

\title{The Opacity of the Intergalactic Medium Measured Along Quasar Sightlines at $z\sim 6$}

\author{Anna-Christina Eilers}
\affiliation{Max-Planck-Institute for Astronomy, K\"onigstuhl 17, 69117 Heidelberg, Germany}
\affiliation{International Max Planck Research School for Astronomy \& Cosmic Physics at the University of Heidelberg}

\author{Frederick B. Davies}
\affiliation{Physics Department, University of California, Santa Barbara, CA 93106-9530, USA}

\author{Joseph F. Hennawi}
\affiliation{Physics Department, University of California, Santa Barbara, CA 93106-9530, USA}
\affiliation{Max-Planck-Institute for Astronomy, K\"onigstuhl 17, 69117 Heidelberg, Germany}

\correspondingauthor{Anna-Christina Eilers}
\email{eilers@mpia.de}

\begin{abstract}
We publicly release a new sample of $34$ medium resolution quasar   spectra at $5.77\leq z_{\rm em}\leq6.54$ observed with the   Echellette Spectrograph and Imager (ESI) on the Keck telescope.   This quasar sample represents an ideal laboratory to study the   intergalactic medium (IGM) during the end stages of the epoch of   reionization, and constrain the timing and morphology of the phase   transition. For a subset of $23$ of our highest signal-to-noise   ratio spectra (S/N$>7$, per $10\,{\rm km\,s^{-1}}$ pixel), we present a new measurement of the Lyman-$\alpha$ (Ly$\alpha$) forest opacity spanning the redshift range $4.8\lesssim z\lesssim6.3$. We   carefully eliminate spectral regions that could be causing biases in   our measurements due to additional transmitted flux in the proximity   zone of the quasars, or extra absorption caused by strong   intervening absorption systems along the line of sight. We compare   the observed evolution of the IGM opacity with redshift to   predictions from a hydrodynamical simulation with uniform   ultraviolet background (UVB) radiation, as well as two semi-numerical patchy reionization models, one with a fluctuating UVB and another with a fluctuating temperature field. Our measurements show a steep rise in opacity at $z\gtrsim5.0$ and an increased scatter and thus support the picture of a spatially inhomogeneous reionization process, consistent with previous work. However, we measure significantly higher optical depths at $5.3\lesssim z\lesssim5.7$ than previous studies, which reduces the contrast between the highest opacity Gunn-Peterson troughs and the average opacity trend of the IGM, which may relieve some of the previously noted tension between these measurements and reionization models.
\end{abstract}

\keywords{intergalactic medium --- epoch of reionization, dark ages --- methods: data analysis --- quasars: absorption lines} 

\section{Introduction}

Determining when and how the epoch of reionization proceeded is one of the major goals of observational cosmology today. During this early evolutionary phase of our universe, the cosmic ``dark ages" following recombination ended, and the intergalactic medium (IGM) transitioned from a neutral state into the ionized medium that we observe today due to the ultraviolet radiation of the first stars, galaxies and quasars. The details of the reionization process not only reflect the nature of these primordial objects, but also the formation of large-scale structure and are therefore a subject of major interest. Despite much progress in the last decade, there are still crucial yet unanswered questions regarding the exact timing and morphology of reionization. 

The most compelling constraints to date on the end of the reionization epoch come from the evolution of the Lyman-alpha (Ly$\alpha$) forest opacity, observed in the spectra of $z\gtrsim6$ quasars \citep{Fan2006, Becker2015, Bosman2018}. 
The detection of transmitted flux spikes in the high redshift quasar spectra and the absence of large Gunn-Peterson (GP) troughs \citep{GunnPeterson1965} below $z\lesssim 5.0$ indicate that the epoch of reionization must be completed by that time. 
Both the steep rise in the observed opacity around $z\gtrsim 5.5$ as well as the increased scatter of the measurements suggest a qualitative change in the ionization state of the IGM \citep{Becker2015}, provoked by a decrease in the ionizing ultraviolet background (UVB) radiation \citep{Calverley2011, WyitheBolton2011}. 

The inferred rapid decline of the UVB radiation has been interpreted as an indication for the end stages of reionization \citep[e.g.][]{Fan2006, BoltonHaehnelt2007c, Calverley2011}. 
\citet{Fan2006} argue further that the increased scatter in the opacity measurements around $z\gtrsim 5.5$ could be explained by strong variations in the UVB radiation field as expected in patchy and inhomogeneous reionization scenarios. 

However, \citet{Lidz2006} argued that large scale density fluctuations alone could explain the significant variations between sightlines. They calculated the scatter arising solely from density fluctuations while assuming a uniform UVB, which gives results comparable to the observations by \citet{Fan2006}. If this was correct, the evidence for patchy reionization based on the observations of the mean opacity would be significantly weakened. 

Measurements of the Ly$\alpha$ forest opacity along additional quasar sightlines by \citet{Becker2015} finally showed that the observed scatter in the optical depth measurements significantly exceeds not only fluctuations expected from the density field alone, but also fluctuating UVB models with a spatially-uniform mean free path of ionizing photons. They posited that, if a fluctuating UVB was in fact the source of the large scatter in optical depth, the mean free path must be spatially variable, supporting the interpretation of probing the end stages of an inhomogeneous hydrogen reionization period. 
Subsequently, \citet{DaviesFurlanetto2016} modeled the UVB with a spatially-varying mean free path and found that the additional fluctuations were sufficient to explain the extra scatter in the optical depth measurements \citep[see also][]{Chardin2017}. 

An alternative explanation for the observed scatter in the mean opacity was presented by \citet{Daloisio2015}, who showed that residual spatial fluctuations in the temperature field could result in extended opacity variations. The spatially varying temperature field is a natural consequence of an extended an inhomogeneous reionization process, wherein regions that reionized early have had time to cool, while regions that reionized late are still hot. The amplitude of the resulting opacity variations then depends directly on the timing and duration of the reionization process.

The largest sample of quasar sightlines to date used for IGM opacity measurements was recently presented by \citet{Bosman2018}. They compare their findings to IGM models that include either a fluctuating UVB or temperature fluctuations and conclude that neither fully captures the observed scatter in IGM opacity. 

In this paper we present a new data set of $34$ high redshift quasar
spectra at $5.77\leq z_{\rm em} \leq 6.54$, which we make publicly
available, and which presents an ideal laboratory for studying the
epoch of reionization and setting constraints on the timing and the
morphology of the reionization process.  For a subset of $23$ quasar
sightlines we present new measurements of the evolution of the mean
opacity of the IGM within the \lya forest between $4.8\lesssim
z\lesssim 6.3$. 
We carefully mask all spectral regions that could be biasing our measurements -- the
region in the immediate vicinity of the quasar, its so-called
proximity zone where the transmitted flux is enhanced due to the
radiation from the quasar itself, as well as all patches exhibiting
additional absorption due to intervening absorption systems such as
damped \lya absorbers (DLAs), along the line of sight to the
quasars. Additionally we correct for possible offsets in the
zero-level of the quasar spectra due to potential systematics in the
data reduction procedure.

We then compare our opacity measurements to predictions from a  hydrodynamical simulation for three different cases --- assuming a uniform UVB radiation field, a fluctuating UVB field, and a fluctuating temperature field --- in order to describe the observed evolution in the IGM opacity, and to assess the excess of inhomogeneities in the density, radiation or temperature field that would be required to explain our measurements. 

This paper is organized as follows: in \S~\ref{sec:data} we describe
our data set and its properties. In this section we also present the data release of this quasar sample within the \texttt{igmspec}\footnote{\url{http://specdb.readthedocs.io/en/latest/igmspec.html}} database. 
The methods we use to continuum normalize the quasar spectra and measure the IGM opacity are described in \S~\ref{sec:methods}. 
The results of the opacity measurements and their evolution with redshift are presented in \S~\ref{sec:results}. 
We compare our measurements to different outputs from a hydrodynamical simulation, which is described in \S~\ref{sec:sims}. 
The implications for the epoch of reionization are discussed in \S~\ref{sec:discussion}, before we conclude in \S~\ref{sec:summary} with a summary of the main results. 
Throughout the paper we assume a cosmology of $h=0.685$, $\Omega_{\rm m}=0.3$ and $\Omega_{\Lambda}=0.7$, which is consistent within the $1\sigma$ errorbars with \citet{Planck2015}.

\section{High Redshift Quasar Sample}\label{sec:data} 

In this section we describe the properties of the data sample of
quasar spectra that we use for our analysis of the IGM opacity. This data set has been previously introduced in \citet{Eilers2017a}, in which we conducted a detailed analysis of the proximity zones of the quasars. We
briefly summarize the details of the observations (\S~\ref{sec:obs}), and
the data reduction procedure (\S~\ref{sec:reduction}). The properties
of this quasar sample are described in \S~\ref{sec:prop}, before presenting the data release at the end of the section (\S~\ref{sec:data_release}).

\subsection{Properties of the Data Set} \label{sec:obs}

Our complete data set consists of $34$ 
quasar spectra at redshifts $5.77\leq z_{\rm em} \leq 6.54$, 
observed at optical wavelengths ($4000$~{\AA} - $10000$~{\AA}) with the Echellette Spectrograph and Imager \citep[ESI;][]{ESI} at the Keck II Telescope in the years 2001 to 2016. 
The data was collected from the Keck Observatory Archive\footnote{\url{https://koa.ipac.caltech.edu/cgi-bin/KOA/nph-KOAlogin}} and complemented with our own observations of four objects ($\rm PSO\,J036+03$, $\rm PSO\,J060+25$, $\rm SDSS\,J0100+2802$, and $\rm SDSS\,J1137+3549$), that we observed in January of 2016. 
All observations were obtained using slit widths ranging from $0.75"-1.0"$, resulting in a resolution of $R\approx 4000-5400$. 
The total exposure times vary from $0.3$~h $\lesssim t_{\rm exp}\lesssim 25$~h resulting in median signal-to-noise ratios in the quasar continuum, at rest-frame wavelengths of $1250${\AA}\,-$1280${\AA}\, between $2\lesssim \rm S/N \lesssim 112$ per pixel. The details of the individual observations are shown in Table~$1$ of \citet{Eilers2017a}.

\subsection{Data Reduction}\label{sec:reduction}

A detailed description of the data reduction can be found in \citet{Eilers2017a} and will be summarized here only briefly. 
All spectra were reduced uniformly using the ESIRedux pipeline\footnote{\url{http://www2.keck.hawaii.edu/inst/esi/ESIRedux/}} developed as part of the XIDL\footnote{\url{http://www.ucolick.org/~xavier/IDL/}} suite of astronomical routines in the Interactive Data Language (IDL). 
This pipeline employs standard data reduction techniques comprising the following: images are overscan subtracted, flat fielded using
a normalized flat field image, and then wavelength calibrated by means of a wavelength image constructed
from afternoon arc lamp calibration images. After identifying the objects in the science frames, we subtracted the
background using $B$-spline fits \citep{Kelson2003, Bernstein2015} to the object free regions of the slit. The profiles of the science objects were also fit with $B$-splines, and an optimal extraction 
was performed on the sky-subtracted frames. We combined one-dimensional spectra of overlapping echelle orders to produce a spectrum for each
exposure, and co-added individual exposures into a final one-dimensional spectrum. For a more detailed description of the applied algorithms, see \citep{Bochanski2009}. 
We further optimized the XIDL ESI pipeline to improve the data
reduction for quasars at high redshift by differentiating two
images (ideally taken during the same observing run) with similar exposure times, analogous to the standard difference imaging techniques performed for near-infrared observations, in order to improve the sky-subtraction especially in the reddest echelle orders, which are affected by fringing. 
This procedure requires dithered exposures for which the trace of the science object lands at different spatial locations on the slit,
Dithered exposures have the additional advantage that different parts of the fringing pattern are being sampled, and hence a combination of different exposures further reduces fringing issues in the data. 
However, since not every observer dithered their object along the slit it was not possible for us to apply this procedure to $\approx 10\%$ of the exposures that we took from the archive. 

We co-added exposures from different observing runs taken by various PIs to maximize the signal-to-noise ratio. 
To combine the data from different observing runs, we weighted each one-dimensional spectrum
by its squared signal-to-noise ratio ($\rm S/N^2$), determined in the quasar continuum region of each spectrum, i.e. at wavelengths longer than the \lya emission line. This ensures that
spectral regions with low or no transmitted flux, which are common in
high redshift quasar spectra, obtain the same weight as regions with more transmitted flux.

All final reduced quasar spectra are shown in Fig.~\ref{fig:spectra1} and \ref{fig:spectra2}, sorted by their emission redshift. 

\startlongtable
\begin{deluxetable*}{lllLLcRc}
\tablewidth{\textwidth}
\tablecaption{Overview of our data sample. \label{tab:overview_data}}
\tablehead{\colhead{object} & \colhead{RAhms} & \colhead{DECdms} & \dcolhead{z_{\rm em}} & \dcolhead{M_{\rm 1450}}  & \colhead{$R_p$ [pMpc]} & \colhead{S/N\tablenotemark{a}} & \colhead{opacity?}}
\startdata
SDSS\,J0002+2550 & $00^\mathrm{h}02^\mathrm{m}39\fs39$ & $+25\degr50\arcmin34\farcs96$ & 5.82 & -27.31 & $5.43\pm 1.49$ & 59 & yes\\
SDSS\,J0005-0006 & $00^\mathrm{h}05^\mathrm{m}52\fs34$ & $-00\degr06\arcmin55\farcs80$ & 5.844 & -25.73 & $2.87\pm 0.40$ & 13 & yes\\
CFHQS\,J0050+3445 & $00^\mathrm{h}55^\mathrm{m}02\fs91$ & $+34\degr45\arcmin21\farcs65$ & 6.253 & -26.7 & $4.09\pm 0.37$ & 18 & yes\\
SDSS\,J0100+2802 & $01^\mathrm{h}00^\mathrm{m}13\fs02$ & $+28\degr02\arcmin25\farcs92$ & 6.3258 & -29.14 & $7.12\pm 0.13$ & 39 & yes\\
ULAS\,J0148+0600 & $01^\mathrm{h}48^\mathrm{m}37\fs64$ & $+06\degr00\arcmin20\farcs06$ & 5.98 & -27.39 & $6.03\pm 0.39$ & 27 & yes\\
ULAS\,J0203+0012 & $02^\mathrm{h}03^\mathrm{m}32\fs38$ & $+00\degr12\arcmin29\farcs27$ & 5.72 & -26.26 & -- & 9 & no\\
CFHQS\,J0210-0456 & $02^\mathrm{h}10^\mathrm{m}13\fs19$ & $-04\degr56\arcmin20\farcs90$ & 6.4323 & -24.53 & $1.32\pm 0.13$ & 1 & no\\
PSO\,J036+03 & $02^\mathrm{h}26^\mathrm{m}01\fs87$ & $+03\degr02\arcmin59\farcs42$ & 6.5412 & -27.33 & $3.64\pm 0.13$ & 11 & yes\\
CFHQS\,J0227-0605 & $02^\mathrm{h}27^\mathrm{m}43\fs29$ & $-06\degr05\arcmin30\farcs20$ & 6.20 & -25.28 & $1.60\pm 1.37$ & 3 & no\\
SDSS\,J0303-0019 & $03^\mathrm{h}03^\mathrm{m}31\fs40$ & $-00\degr19\arcmin12\farcs90$ & 6.078 & -25.56 & $2.21\pm 0.38$ & 2 & no\\
SDSS\,J0353+0104 & $03^\mathrm{h}53^\mathrm{m}49\fs73$ & $+01\degr04\arcmin04\farcs66$ & 6.072 & -26.43 & -- & 13 & no\\
PSO\,J060+25 & $04^\mathrm{h}02^\mathrm{m}12\fs69$ & $+24\degr51\arcmin24\farcs43$ & 6.18 & -26.95 & $4.17\pm 1.38$ & 12 & yes\\
SDSS\,J0818+1722 & $08^\mathrm{h}18^\mathrm{m}27\fs40$ & $+17\degr22\arcmin52\farcs01$ & 6.02 & -27.52 & $5.89\pm 1.42$ & 7 & yes\\
SDSS\,J0836+0054 & $08^\mathrm{h}36^\mathrm{m}43\fs86$ & $+00\degr54\arcmin53\farcs26$ & 5.81 & -27.75 & $5.06\pm 0.40$ & 112 & yes\\
SDSS\,J0840+5624 & $08^\mathrm{h}40^\mathrm{m}35\fs30$ & $+56\degr24\arcmin20\farcs22$ & 5.8441 & -27.24 & $7.39\pm0.30$\tablenotemark{b} & 26& yes \\
SDSS\,J0842+1218 & $08^\mathrm{h}42^\mathrm{m}29\fs43$ & $+12\degr18\arcmin50\farcs58$ & 6.069 & -26.91 & $6.47\pm 0.38$ & 10 & yes\\
SDSS\,J0927+2001 & $09^\mathrm{h}27^\mathrm{m}21\fs82$ & $+20\degr01\arcmin23\farcs64$ & 5.7722 & -26.76 & $4.68\pm 0.15$ & 7 & no\\
SDSS\,J1030+0524 & $10^\mathrm{h}30^\mathrm{m}27\fs11$ & $+05\degr24\arcmin55\farcs06$ & 6.309 & -26.99 & $5.93\pm 0.36$ & 24 & yes\\
SDSS\,J1048+4637 & $10^\mathrm{h}48^\mathrm{m}45\fs07$ & $+46\degr37\arcmin18\farcs55$ & 6.2284 & -27.24 & -- & 43 & no\\
SDSS\,J1137+3549 & $11^\mathrm{h}37^\mathrm{m}17\fs73$ & $+35\degr49\arcmin56\farcs85$ & 6.03 & -27.36 & $6.98\pm 1.42$ & 22 & yes\\
SDSS\,J1148+5251 & $11^\mathrm{h}48^\mathrm{m}16\fs65$ & $+52\degr51\arcmin50\farcs39$ & 6.4189 & -27.62 & $4.58\pm 0.13$ & 35 & yes\\
SDSS\,J1250+3130 & $12^\mathrm{h}50^\mathrm{m}51\fs93$ & $+31\degr30\arcmin21\farcs90$ & 6.15 & -26.53 & $6.59\pm 1.38$ & 8 & yes \\
SDSS\,J1306+0356 & $13^\mathrm{h}06^\mathrm{m}08\fs27$ & $+03\degr59\arcmin26\farcs36$ & 6.016 & -26.81 & $5.39\pm 0.38$ & 37 & yes\\
ULAS\,J1319+0950 & $13^\mathrm{h}19^\mathrm{m}11\fs30$ & $+09\degr50\arcmin51\farcs52$ & 6.133 & -27.05 & $3.84\pm 0.14$ & 10 & yes\\
SDSS\,J1335+3533 & $13^\mathrm{h}35^\mathrm{m}50\fs81$ & $+35\degr33\arcmin15\farcs82$ & 5.9012 & -26.67 & $0.78\pm 0.15$ & 6 & no\\
SDSS\,J1411+1217 & $14^\mathrm{h}11^\mathrm{m}11\fs29$ & $+12\degr17\arcmin37\farcs28$ & 5.904 & -26.69 & $4.60\pm 0.39$ & 29 & yes\\
SDSS\,J1602+4228 & $16^\mathrm{h}02^\mathrm{m}53\fs98$ & $+42\degr28\arcmin24\farcs94$ & 6.09 & -26.94 & $7.11\pm 1.40$ & 21 & yes\\
SDSS\,J1623+3112 & $16^\mathrm{h}23^\mathrm{m}31\fs81$ & $+31\degr12\arcmin00\farcs53$ & 6.2572 & -26.55 & $5.05\pm 0.14$ & 9 & yes\\
SDSS\,J1630+4012 & $16^\mathrm{h}30^\mathrm{m}33\fs90$ & $+40\degr12\arcmin09\farcs69$ & 6.065 & -26.19 & $4.80\pm 0.38$ & 15 & yes\\
CFHQS\,J1641+3755 & $16^\mathrm{h}41^\mathrm{m}21\fs73$ & $+37\degr55\arcmin20\farcs15$ & 6.047 & -25.67 & $3.98\pm 0.38$ & 3 & no\\
SDSS\,J2054-0005 & $20^\mathrm{h}54^\mathrm{m}06\fs49$ & $-00\degr05\arcmin14\farcs80$ & 6.0391 & -26.21 & $3.17\pm 0.14$ & 15 & yes\\
CFHQS\,J2229+1457 & $22^\mathrm{h}29^\mathrm{m}01\fs65$ & $+14\degr57\arcmin09\farcs00$ & 6.1517 & -24.78 & $0.45\pm 0.14$ & 2 & no\\
SDSS\,J2315-0023 & $23^\mathrm{h}15^\mathrm{m}46\fs57$ & $-00\degr23\arcmin58\farcs10$ & 6.117 & -25.66 & $3.70\pm 1.39$ & 14 & yes\\
CFHQS\,J2329-0301 & $23^\mathrm{h}29^\mathrm{m}08\fs28$ & $-03\degr01\arcmin58\farcs80$ & 6.417 & -25.25 & $2.45\pm 0.35$ & 2 & no\\
\enddata
\tablecomments{The columns show the object name, the coordinates of the quasar given in $\rm RAhms$ and $\rm DECdms$, the emission redshift and 
the quasar's magnitude $M_{\rm 1450}$, the measurements for their  proximity zones, and the $\rm S/N$ ratio of the quasar spectrum. The last column states, whether we used the quasar sightline for the IGM opacity analysis in this paper. }
\tablenotetext{a}{Median $\rm S/N$ per $10\,\rm km\,s^{-1}$ pixel; estimated between $1250$~{\AA}$\leq\lambda_{\rm rest}\leq1280$~{\AA}. }
\tablenotetext{b}{Estimate of the proximity zone conservatively assuming $\Delta v=5000$~\kms, because the proximity zone is prematurely truncated due to associated absorption systems \citep[see Appendix A in][]{Eilers2017a}. }
\end{deluxetable*}

\subsection{Quasar Properties}\label{sec:prop}

The properties of all quasars in our data sample, such as the emission redshift $z_{\rm em}$, the absolute magnitude $M_{\rm 1450}$ at $\lambda_{\rm rest} = 1450$~{\AA} in the rest frame, the size of the proximity zone $R_p$ of the quasars and the S/N ratio of their spectra, are presented in Table~\ref{tab:overview_data}. 

The proximity zone measurements are taken from \citet{Eilers2017a}. The uncertainties of these measurements arise solely from uncertainties in the redshift estimate, since these errors provide the largest source of uncertainty for the proximity zone measurements. 
For one object, $\rm SDSS~J0840+5624$ we do not report a measurement of its proximity zone, because associated absorption systems located in the immediate vicinity of the quasar, prematurely truncate its proximity zone \citep[see Appendix A in][]{Eilers2017a}. 
Thus we assume conservatively a region of $\Delta v=5000$~\kms\, to be within the influence of the quasar's radiation, resulting in an effective $R_{\rm p} = 7.39\pm0.30$. 

For three objects ($\rm ULAS~J0203+0012$, $\rm SDSS~J0353+0104$, and $\rm SDSS~J1048+4637$) no estimates of their proximity zones have been determined. These quasars show broad absorption line (BAL) features, which make a precise and unbiased estimate of their proximity zones unfeasible. Due to the additional absorption features in their spectra we exclude these objects from the analysis of the IGM opacity.

\begin{figure*}[ht!]
\centering
\includegraphics[width=\textwidth]{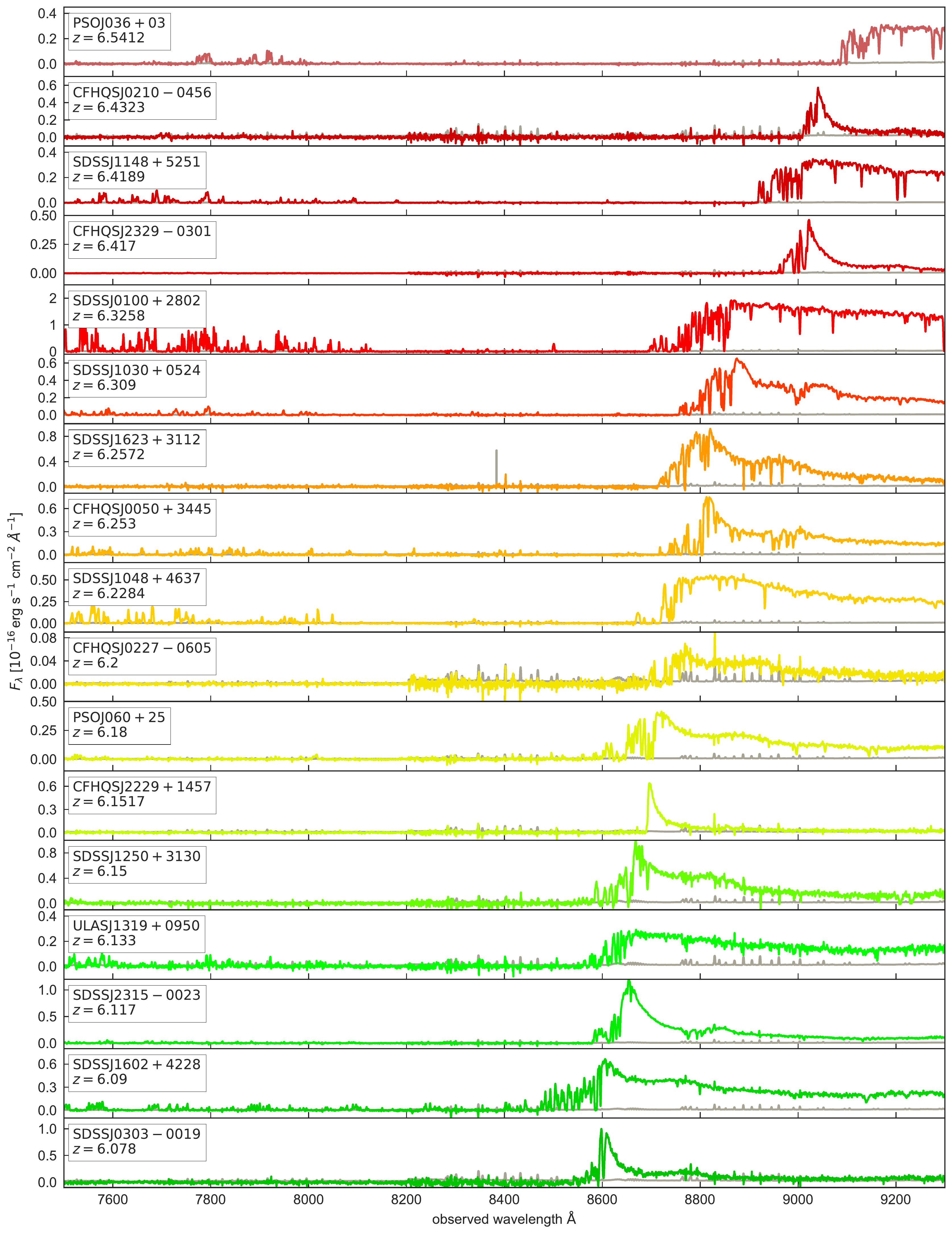}
\caption{Spectra of all quasars in our data sample sorted by redshift. \label{fig:spectra1}} 
\end{figure*}

\begin{figure*}[ht!]
\centering
\includegraphics[width=\textwidth]{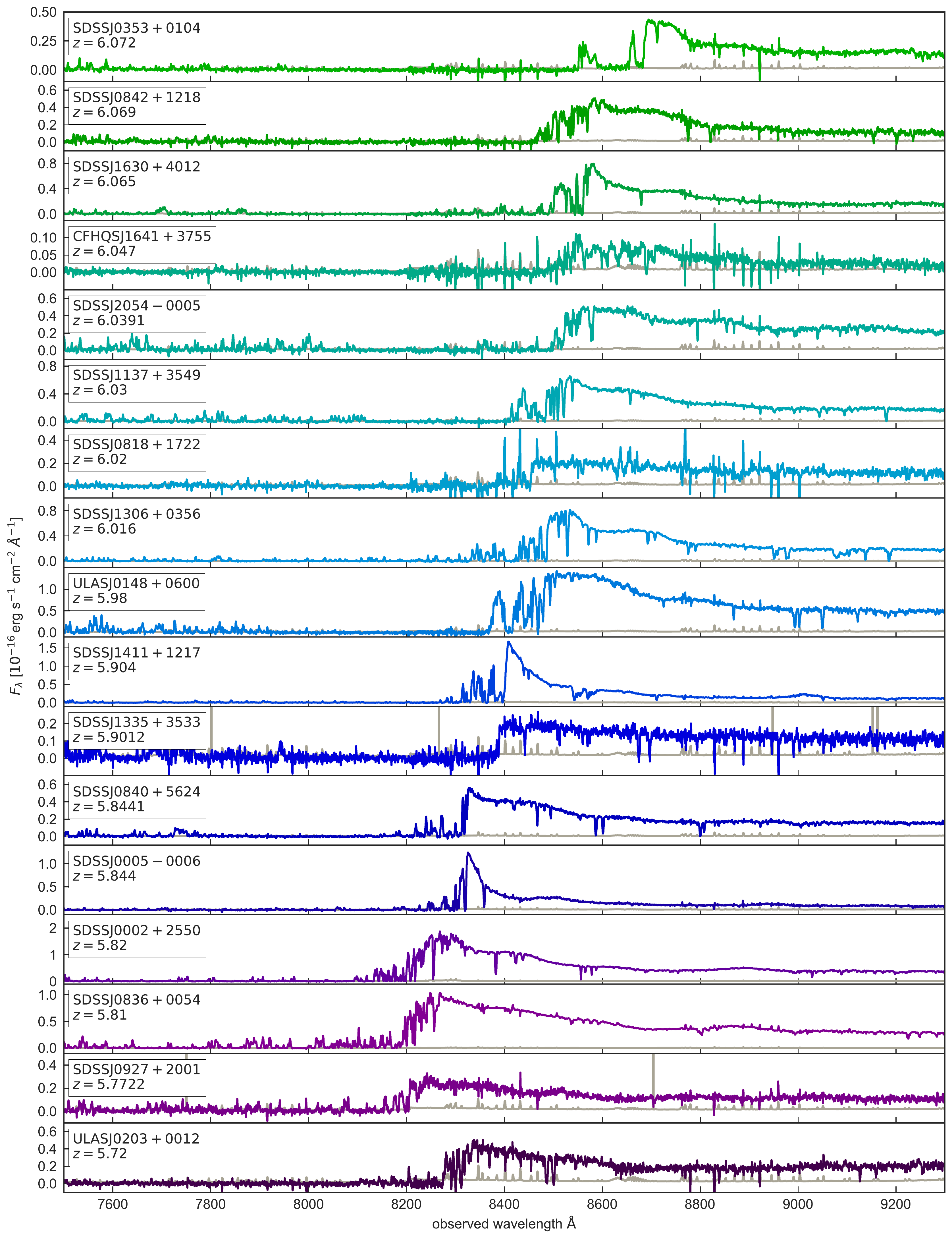}
\caption{Continuation of Fig.~\ref{fig:spectra1}. \label{fig:spectra2}} 
\end{figure*}

\subsection{Data Release}\label{sec:data_release}

Our new data set including the final co-added spectra and their noise vectors together with the estimated quasar continua (see \S~\ref{sec:cont}) will be available via the \texttt{igmspec} database\footnote{\url{http://specdb.readthedocs.io/en/latest/igmspec.html}} \citep{Prochaska2017}, as well as additional meta data on the quasars. A catalog for the data release comprising the main properties of the data set is shown in Tab.~\ref{tab:catalog} in Appendix~\ref{sec:catalog}. 

\section{Methods}\label{sec:methods}

In order to measure the IGM opacity the quasar spectra must be normalized to their unabsorbed continua. In this section we explain our method for continuum normalizing the quasar spectra (\S~\ref{sec:cont}) and present the details of the IGM opacity measurements (\S~\ref{sec:measurements}). In the last part of this section (\S~\ref{sec:offset}) we describe a procedure for correcting small offsets in the zero-level of the quasar spectra. 

\subsection{Quasar Continuum Normalization}\label{sec:cont}

\begin{figure*}[!t]
\centering
\includegraphics[width=.95\textwidth]{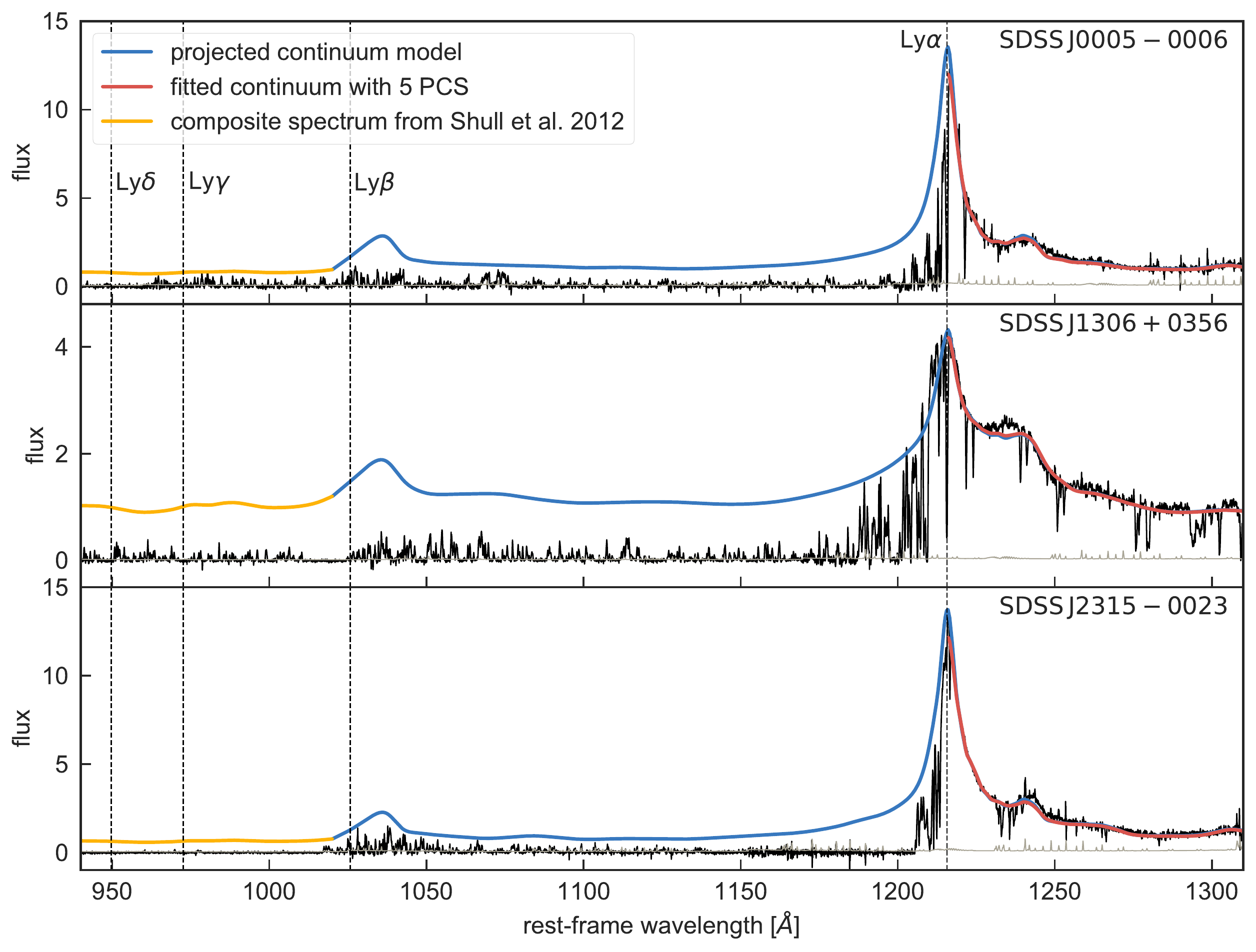}
\caption{Examples of three quasar spectra in our data sample and their continuum models. The continua are first fitted with five PCS from \citet{Paris2011} to the unabsorbed quasar continuum at wavelength $\lambda_{\rm rest}\geq 1215.67$~{\AA} (red part) and afterwards projected onto the blue side of the quasar spectrum (blue part). The continuum model is then augmented to lower wavelengths by appending the composite spectrum from \citet{Shull2012} at $\lambda_{\rm rest}\leq 1020$~{\AA} (yellow part). 
The vertical dashed line marks the location of the Ly$\alpha$, Ly$\beta$, Ly$\gamma$, Ly$\delta$ emission line. \label{fig:cont}} 
\end{figure*}

We fit the quasar continua and normalize the spectra in a similar manner as previously conducted by \citet{Eilers2017a}. We will summarize the main steps here briefly, but refer the reader to the previous paper for more details. 

All spectra were normalized to unity at $\lambda_{\rm rest}=1280$~{\AA} in a spectral region that is free of emission lines. The quasar continuum was then estimated with principal component spectra (PCS) from a principal component analysis (PCA) of lower redshift quasar spectra \citep{Paris2011}. The concept of PCA is to represent each continuum spectrum $\ket{q_{\lambda}}$ at wavelength $\lambda$ by a reconstructed spectrum comprising a mean spectrum $\ket{\mu_{\lambda}}$ and a sum of $m$ weighted PCS $\ket{\xi_{\lambda}}$, i.e. 
\begin{equation}
\ket{q_{\lambda}}\approx\ket{\mu_{\lambda}} + \sum_{i=1}^m \alpha_{i}\ket{\xi_{i, \lambda}},\label{eq:pca}
\end{equation}
where the index $i$ refers to the $i$th PCS and $\alpha_{i}$ denotes its weight. 

Since the quasars in our data sample are all at $z_{\rm em}\sim6$, they experience substantial absorption due to intervening neutral hydrogen within the IGM bluewards of the \lya emission line. Thus the continuum estimate was performed solely on the unabsorbed continuum spectrum redwards of the \lya emission line with a set of PCS from \citet{Paris2011} covering the wavelengths $1215.67$~{\AA}$\leq\lambda_{\rm rest}\leq 2000$~{\AA}. We take the model that minimizes $\chi^2$ using the noise vector from the spectra as the best estimate.

 In order to obtain coefficients $\bm{\alpha}$ for a set of PCS that
 cover the \textit{entire} spectral region between
 $1020$~{\AA}~$\leq\lambda_{\rm rest}\leq 2000$~{\AA}, we use a
 projection matrix $\bm{P}$ to project the estimated coefficients for
 the PCS \textit{redwards} of Ly$\alpha$ onto this new set of
 coefficients for the entire spectrum.  The projection matrix $\bm{P}$
 has been computed by \citet{Paris2011} using the set of PCS for
 both the red wavelength side only and the whole spectral region
 covering wavelengths bluewards and redwards of \lya. Hence
\begin{equation}
\bm{\alpha} = \bm{P} \cdot \bm{\alpha_{\rm red}}.  
\end{equation}
This new set of coefficients together with eqn.~\ref{eq:pca} provides a continuum model for each quasar covering all wavelengths $1020$~{\AA}~$\leq\lambda_{\rm rest}\leq 2000$~{\AA}. \citet{Paris2011} estimate that the median uncertainty on the transmitted flux in the \lya forest to be $|\Delta F_{\rm forest}| \approx 5\%$. However, since we do not take all PCA components into account and do not have the full spectral coverage up to $\lambda_{\rm rest} = 2000$~{\AA} to estimate the continua, the uncertainty on the continua in our quasar spectra is most likely higher, i.e. $|\Delta F_{\rm forest}| \approx 10-20\%$. 


For an estimate of the continua at lower wavelength we take the composite quasar spectrum provided by \citet{Shull2012}, constructed from $22$ low redshift quasars observed with the Cosmic Origins Spectrograph (COS) on the HST that extends from $550$~{\AA} to $1750$~{\AA} in the rest frame, and re-scale the composite spectrum to match the PCA constructed continuum model at $\lambda_{\rm rest}=1020$~{\AA}. We augment the continuum model by simply appending the composite spectrum at wavelengths $\lambda_{\rm rest}<1020$~{\AA}. Note however, that we do not use the spectrum at $\lambda_{\rm rest}<1020$~{\AA} for the analysis of the \lya forest opacity. A few example quasar spectra from our data set and its continuum model are shown in Fig.~\ref{fig:cont}, all remaining quasar spectra that we analyzed and their estimated continua are shown in Fig.~\ref{fig:spectra_all_cont} in Appendix~\ref{sec:all_cont_fits}. 

Note that in some cases the \ion{N}{5} line at $\lambda_{\rm rest} = 1240\,\rm{\AA}$ is not very well represented by the continuum fit. This behavior occurs when the \ion{N}{5} line is slightly blue- or redshifted compared to the systemic redshift of the quasar and those shifts are not accounted for in the PCA basis. Additionally, the continuum for quasar spectra with very weak emission lines, such as $\rm SDSS\,J0100+2802$ or $\rm SDSS\,J1148+5251$, is not very well captured by the PCA. 

\subsection{Measuring the Optical Depth of the IGM}\label{sec:measurements}

We exclude BAL quasars ($\rm ULAS~J0203+0012$, $\rm SDSS~J0353+0104$, and $\rm SDSS~J1048+4637$) from our IGM optical depth sample to avoid potential contamination by broad non-IGM absorption in the Ly$\alpha$ forest, and quasars with only very low S/N data, i.e. $\rm S/N < 7$, 
 whose spectra are more subject to systematic errors. 
 Our final IGM optical depth sample then consists of $23$ quasar spectra out of the original $34$. 

We estimate the mean opacity of the IGM by means of the effective optical depth, which is defined as 
\begin{equation}
\tau_{\rm eff} = -\ln\,\langle F\rangle, \label{eq:tau} 
\end{equation}
where $F$ is the continuum normalized flux. The effective optical depth is measured in discrete spectral bins along the line of sight of each quasar. We chose fixed comoving bins of size $50$~comoving Mpc$\,h^{-1}$ (\cmpch) \citep{Becker2015}, which contains a similar path length as the bins of size  $\Delta z=0.15$ at $z\sim5-6$ previously applied by \citet{Fan2006}. 

In order to avoid biases in the measurement of the opacity of the IGM, we mask the spectral region around each quasar that is strongly influenced and ionized by the quasar's own radiation. We use the measurements for the proximity zones $R_p$ as an estimate the influenced region. 
 However, the proximity zone is defined such that it does not completely reach out to the ionization front expanding from the quasar. 
Thus the influence of the quasar's radiation is expected to be still present outside of its measured proximity zone \citep{Eilers2017a}, since the radiation of the quasar still dominates the UVB radiation, i.e. $\Gamma_{\rm QSO} \gg \Gamma_{\rm UVB}$ at $R_{\rm p}$. 
Hence, we mask an additional $2.5$~proper Mpc (pMpc) around each quasar, i.e. the masked region measures $R_{\rm p}+2.5$~pMpc, in order to eliminate all enhanced transmission due to the quasar's radiation. 

Thus we choose the maximum wavelength that we consider for opacity measurements to lie just blueward of this masked region. 
The minimum wavelength we consider for measurements within the \lya 
forest is $1030$~{\AA} 
in the rest frame, in order to account for possible redshift uncertainties.  

\begin{figure*}[!t]
\includegraphics[width=\textwidth]{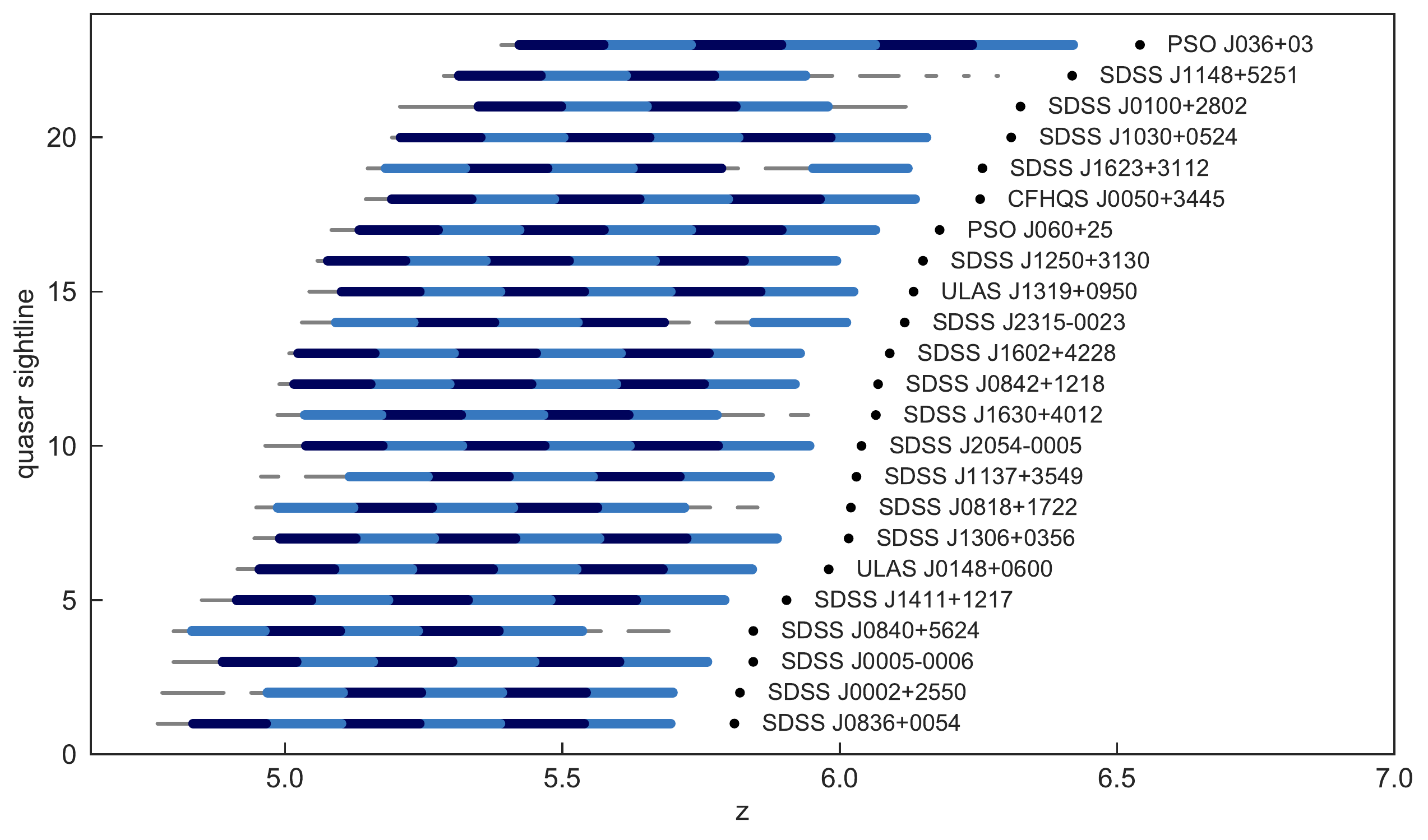}
\caption{Redshift coverage of each quasar spectrum used for the opacity measurements of the IGM. The different blue colored regions show the $50$~\cmpch\, bins within the \lya forest. 
The remaining unused coverage of each spectrum is shown in grey. \label{fig:coverage}} 
\end{figure*}

Another possible contamination in the measurement of optical depths are intervening absorbers along the line of sight, such as damped \lya absorption systems (DLAs) or other low-ion metal absorbers which are likely associated with relatively high \ion{H}{1} column density ($N_{\rm HI}\ga10^{19}$ cm$^{-2}$). We mask the regions in the quasar spectra 
around these absorbers, as they reflect an absorption signature that is not typically resolved in IGM simulations. 
To this end, we searched for low-ion metal absorption lines, such as e.g. \ion{Al}{2}, \ion{Fe}{2}, and \ion{O}{1}, associated with absorbers in the continuum spectra redwards of the Ly$\alpha$ emission line that are located at the same redshift as a spectral region showing complete absorption in the 
forest of the spectrum. 
Additionally, we searched through the literature for DLAs and low-ion metal absorbers along the quasar sight lines in our sample. For each absorber, we conservatively masked the spectral region around the absorption system within $\pm 30$~{\AA} in the observed wavelength frame in the \lya 
forest at the corresponding wavelength.
Spectral bins containing absorbers were then excluded from the IGM opacity measurements.
Table~\ref{tab:dla} shows a compilation of all identified absorbers along the line of sight to the quasars in our sample. Note that most of the identified absorbers have been already found by other authors, since most quasars in our data set have been previously known and been observed. 

Additionally we mask all spuriously high pixels within the \lya 
forest of the quasar spectra, by checking for single pixels showing $F>1$ in the continuum normalized spectra. 
Because sky-subtraction systematics occasionally result in large
negative sky-subtraction residuals, we also mask all negative flux
pixels with the $2.5\%$ lowest S/N, to avoid biases due to large
uncertainties in pixels that fall onto sky lines and have large negative
residuals. 

After masking all low-ion metal absorption systems, proximity zones, and spuriously high and negative pixels, the combined usable path length for the opacity measurements is $6350$~\cmpch for the $23$ quasar sight lines in our data sample. The spectral regions in which we measure the mean flux and calculate its effective optical depth are shown for each quasar as the dark and light blue colored bars 
in Fig.~\ref{fig:coverage}. Masked regions are shown in white. The
grey regions show pathlength that are in principle usable but are not
used, because the remaning unmaksed region would be smaller than our
our chosen bin size of $50$~\cmpch.

\begin{deluxetable}{lLLL}
\tablecaption{Intervening low-ion absorption systems along the line of sight. \label{tab:dla}}
\tablehead{\colhead{object} & \colhead{$z_{\rm em}$} & \colhead{$z_{\rm abs}$} & \colhead{Ref.\tablenotemark{a}}}
\startdata
SDSS~J1148+5251 & 6.4189 & 6.0115 & 1 \\[-2 pt]
 & & 6.1312 & 1\\[-2 pt]
 & & 6.1988 & 1\\[-2 pt]
 & & 6.2575 & 1\\
SDSS~J2054-0005 & 6.0391 & 5.9776 & 1 \\
SDSS~J2315-0023 & 6.117 & 5.7529 & 1 \\
SDSS~J1630+4012	& 6.065 & 5.8865 & 1 \\
SDSS~J1137+3549 & 6.03 & 5.0124 & 1 \\
SDSS~J1623+3112 & 6.2572 & 5.8415 & 1\\
SDSS~J0840+5624 & 5.8441 & 5.5940 & 2 \\
SDSS~J0002+2550 & 5.82 & 4.914 & 3 \\
SDSS~J0100+2802 & 6.3258 & 6.1437 & 3 \\
SDSS~J0818+1722 & 6.02 & 5.7911 & 1 \\[-2 pt]
 & & 5.8765 & 1 \\
\enddata
\tablecomments{The columns show the name of the object and its emission redshift, the absorption redshift of the intervening low-ion absorber and the reference therefor.}
\tablenotetext{a}{Reference for low-ion absorbers. 1: \citet{Becker2011}, 2: \citet{Ryan-Weber2009}, 3: this work. }
\end{deluxetable}

\subsection{Correcting for Offsets in the Zero-Level of the Spectra}\label{sec:offset}

The noise for pixels with no intrinsic flux should be symmetrically distributed around zero, since pixels with zero flux have equal probability to be scattered into positive or negative values. This idea was applied by \citet{McGreer2011}, for example, to estimate the number of so called ``dark pixels" that have a flux value consistent with zero. 
However, a detailed inspection of the quasar spectra in our data set reveals that the zero-level in some spectral regions can be slightly biased, i.e. we do not observe a symmetric distribution around zero in the noise, possibly caused by sky-subtraction systematics present in a small fraction of exposures. But even tiny offsets in the zero-level can cause large differences in the opacity estimates, especially in regions with very little transmitted flux. Because these are the regions we are particularly interested in, we correct for small offsets in the zero-level of the spectra. These offsets are calculated and applied to each $50\,$\cmpch spectral bin individually in order to avoid correlations between the optical depth measurements. 

A detailed description of this correction procedure can be found in Appendix~\ref{sec:offset_details}. We estimate that the systematic error in the mean flux $\langle F^{\rm Ly\alpha}\rangle$ due to corrections of the zero-level offset is 
\begin{align*}
\sigma_{\langle F^{\rm Ly\alpha}\rangle} \approx 0.0067, 
\end{align*}
i.e. less than $<1\%$, and thus constitutes only a minor correction to the optical depths measurements. 

\begin{figure*}[!t]
\centering
\includegraphics[width=\textwidth]{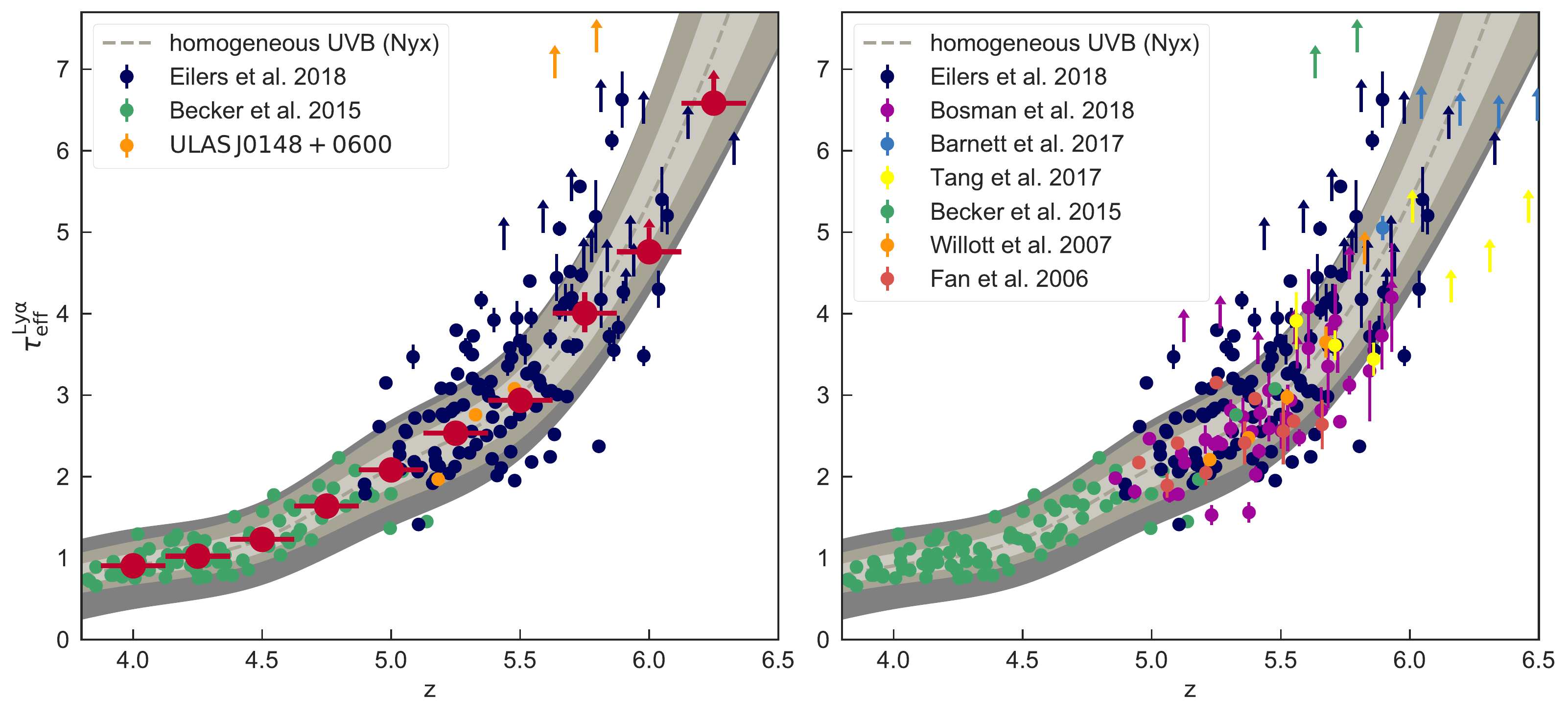}
\caption{Evolution of the \lya forest effective optical depth. \textit{Left panel:} The dark blue data points show our IGM opacity measurements. The green and orange data points are measurements of the optical depth performed by \citet{Becker2015}, orange indicating the measurements of $\rm ULAS\,J0148+0600$. This data set we consider the master compilation sample. The large red data points show the mean redshift evolution averaged over bins of $\Delta z = 0.25$, their uncertainties are determined via bootstrapping. The grey underlying region shows the predicted redshift evolution from radiative transfer simulations assuming a uniform UVB model. We have simulation outputs in steps of $\Delta z = 0.5$ and use a cubic spline function to interpolate the shaded regions between the redshifts of the outputs. The light and medium grey shaded regions indicate the $68$th and $95$th percentile of the scatter expected from density fluctuations in the simulations, whereas the dark grey region shows any additional scatter due to $\sim 20\%$ continuum uncertainties.  \textit{Right panel:} Compilation of all opacity measurements found in the literature along quasar sightlines that are not in our data sample and that have been calculated within similar spectral bin sizes. \label{fig:tau_alpha}} 
\end{figure*}

\section{Results}\label{sec:results}

We compute the effective optical depth $\tau_{\rm eff}$ from the measurements of the observed mean flux $\langle F^{\rm obs}\rangle$ in bins of $50$~cMpc\,$h^{-1}$ using Eqn.~\ref{eq:tau}. We list all measurements of the mean flux within the \lya 
forest for each bin along all $23$ quasar sight lines in our data sample in Tab.~\ref{tab:lya_flux} 
in Appendix~\ref{sec:mean_flux}. All spectral bins indicating the respective measurements of $\langle F^{\rm obs}\rangle$ and $\tau^{\rm Ly\alpha}_{\rm eff}$ are shown in Fig.~\ref{fig:spec_bins_a}, \ref{fig:spec_bins_b}, and \ref{fig:spec_bins_c} in Appendix~\ref{sec:spec_bins}. 
If the mean flux is detected with less than $2\sigma$ significance or if we measure a negative mean flux, we adopt a lower limit on the optical depth at $\tau_{\rm eff} =-\ln(2\,\sigma_{\langle F^{\rm obs}\rangle})$ consistent with previous works. Note that we do not include the systematic uncertainty on the mean flux ($\sim 10-20\%$, see \S~\ref{sec:cont}) arising from the PCA continuum estimate. This uncertainty on the mean flux would lead to an additional  uncertainty on $\tau^{\rm Ly\alpha}_{\rm eff}$ of $\sim 2-5\%$, when most of the flux is absorbed in the quasar spectra \citep[see also Fig.~$7$ in][]{Becker2015}. The results of the optical depth measurements within the \lya forest, plotted as a function of redshift, are shown in Fig.~\ref{fig:tau_alpha}. 

In Fig.~\ref{fig:tau_alpha} 
our new measurements are shown in dark blue in both panels. 
For the majority of quasar spectra that we use in our analysis the IGM opacity has been analyzed before. However, we have co-added the data from multiple observation runs (see \S~\ref{sec:reduction} for details), in order to achieve higher signal-to-noise.  
However, for one object, $\rm ULAS~J0148+0600$, the data obtained by \citet{Becker2015} with VLT/X-Shooter in a $10$~h observation, has a higher $\rm S/N$ ratio than our spectrum. This sightline exhibits a particularly deep GP trough \citep{Becker2015}, and hence the enhanced $\rm S/N$ ratio results in more stringent opacity limits, representing the strongest fluctuations in the IGM opacity at this redshift. In order to model the IGM fluctuations correctly, it is important to include these outliers \citep{Chardin2015, Daloisio2015, DaviesFurlanetto2016}. 
Thus we construct a master compilation of opacity measurements, presented in the left panel of Fig.~\ref{fig:tau_alpha}, and replace our optical depth measurements within the \lya forest along just the sightline of $\rm ULAS~J0148+0600$ with the more precise measurements obtained by \citet{Becker2015} (orange data points). This mainly adds the two most stringent limits at $z=5.634$ and $z=5.796$ to our analysis, since the better data quality results in higher sensitivity in the GP troughs. The lower redshift $\tau_{\rm eff}^{\rm Ly\alpha}$ measurements for this object are consistent with our measurements. This master compilation is shown in the left panel of Fig.~\ref{fig:tau_alpha}. 
Additionally, we also show the lower redshift $\tau_{\rm eff}^{\rm Ly\alpha}$ measurements from \citet{Becker2015} (green data points). 

We present the average opacity evolution by calculating the mean flux and the bootstrapped error on the mean in bins of $\Delta z = 0.25$ 
within the \lya 
forest from the master compilation set, shown in the left panel of Fig.~\ref{fig:tau_alpha}. 
We then compute the binned opacity values via $\tau_{\rm eff} = -\ln\langle F\rangle$, where $\langle F\rangle$ is the mean flux computed in these bins. 
The uncertainties on the opacity values with uncertainties also determined via bootstrapping are
shown as the red data points and tabulated in
Tab.~\ref{tab:comp_lya}. 
Similar to the individual \taueff measurements, we adopt a limit if the mean flux in the bin is measured with less than $2\sigma$ significance (where $\sigma$ is here the bootstrap errors on the mean flux). 

Note that we do not take any systematic errors on the mean flux measurements into account that could, for instance, result from uncertainties in the continuum estimation. The dark grey regions give an estimate of the additional scatter expected due to continuum uncertainties of $\sim 20\%$, which are negligible at high redshift, where the transmitted flux is low and the scatter is dominated by fluctuations along different sightlines \citep{Becker2015, Eilers2017b}.

The right panel of Fig.~\ref{fig:tau_alpha} 
compares our data set to opacity measurements from additional sightlines from the literature that are not in our data sample. 
The additional data points come from the sightlines of $\rm SDSS\,J0144-0125$ and $\rm SDSS\,J1436+5007$ \citep{Fan2006}, $\rm CFHQS\,J1509-1749$ \citep{Willott2007}, 
$\rm ULAS\,J1120+0641$ \citep{Barnett2017}, $\rm PSO\,J006.1240+39.221$ \citep{Tang2017}, and $\rm J0323-4701$, $\rm J0330-4025$, $\rm J0410-4414$, $\rm J0454-4448$,  $\rm J0810+5105$, $\rm J1257+6349$, $\rm J1609+3041$, $\rm J1621+5155$, $\rm J2310+1855$ \citep{Bosman2018}. In most of these analyses the bins were chosen to be $\Delta z = 0.15$, following \citet{Fan2006}. This bin size covers roughly the same spectral region as the chosen bin size of $50$~\cmpch in our analysis and the one by \citet{Becker2015} at $z\sim 6$, but in the redshift interval of $5\lesssim z\lesssim 7$, the bin size changes quite significantly. Overall the agreement between the various measurements with our new analysis is good, but we chose not to add these measurements to the master compilation, because of the different pathlengths used to construct the measurements, very low $\rm S/N$ data or the variety of different instruments and data reduction pipelines used to obtain the spectra, which enlarges the systematic uncertainties on these measurements (see \S~\ref{sec:sys_errors}).

\begin{deluxetable}{LRRRc}
\tablecaption{Measurements of the average flux and optical depth within the \lya forest of our master compilation sample.  \label{tab:comp_lya}}
\tablehead{\dcolhead{z_{\rm abs}} & \dcolhead{\langle F \rangle} & \dcolhead{\sigma_{\langle F\rangle}} & \dcolhead{\langle\tau_{\rm eff}^{\rm Ly\alpha}\rangle} & \dcolhead{\sigma_{\langle\tau_{\rm eff}^{\rm Ly\alpha}\rangle}}}
\startdata
4.0 & 0.4046 & 0.0151 & 0.9049 & 0.0372 \\
4.25 & 0.3595 & 0.0112 & 1.0230 & 0.0311 \\
4.5 & 0.2927 & 0.0190 & 1.2286 & 0.0651 \\
4.75 & 0.1944 & 0.0150 & 1.6381 & 0.0770 \\
5.0 & 0.1247 & 0.0132 & 2.0818 & 0.1060 \\
5.25 & 0.0795 & 0.0078 & 2.5321 & 0.0984 \\
5.5 & 0.0531 & 0.0058 & 2.9357 & 0.1090 \\
5.75 & 0.0182 & 0.0045 & 4.0057 & 0.2469 \\
6.0 & 0.0052 & 0.0043 & >4.7595 & --- \\
6.25\tablenotemark{a} & -0.0025 & 0.0007 & >6.5843 & --- \\
\enddata
\tablecomments{The columns show the mean redshift $z_{\rm abs}$ of the redshift bins of size $\Delta z = 0.25$, the averaged flux $\langle F\rangle$ and its uncertainty $\sigma_{\langle F\rangle}$ determined via bootstrapping, and the mean optical depth $\langle\tau_{\rm eff}^{\rm Ly\alpha}\rangle$ in that redshift bin and its error $\sigma_{\langle\tau_{\rm eff}^{\rm Ly\alpha}\rangle}$, also determined via boostrapping. 
\tablenotetext{a}{Note that this redshfit bin only contains two measurements. }}
\end{deluxetable}

\section{Comparison to Other Studies}\label{sec:sys_errors}

For several quasar sight lines in our data sample the optical depth has been measured previously by \citet{Fan2006} and \citet{Becker2015}, and more recently by \citet{Bosman2018}. However, the quality of the data and the methods to analyze the data differ. 
Here, we carry out a detailed comparison of our methods and measurements to previous work, and discuss potential systematic uncertainties (\S~\ref{sec:comp_fan}) and resulting discrepancies in the cumulative distribution functions (CDFs) of the optical depth (\S~\ref{sec:comp_cdf}).

\subsection{Estimating Systematic Uncertainties}\label{sec:comp_fan}

We compare the measurements of the IGM opacity for the $16$ quasar sightlines that are both part of our analysis and the data set of \citet{Fan2006} and are not BAL quasars. 
The spectra from \citet{Fan2006} partially overlap with our data set, but six quasars
were observed with a different telescope and instrument (MMT -- MMT Red Channel, Hobby-Eberly Telescope (HET), Kitt Peak (KP) -- KP 4m MARS) and eight quasars have additional Keck/ESI data. We have reduced and stacked all of the Keck/ESI observations from the archive and thus the spectra analyzed in this paper have an improved quality due to their longer exposure time. 

Additionally, our methods to analyze the data differ. For instance, \citet{Fan2006} applied a power-law to the red side of the quasar spectra to estimate the quasar continua, whereas we estimated the quasar continua by a PCA (see \S~\ref{sec:cont}). 
In our analysis we mask all spectral regions containing low-ion metal absorption systems, while in previous work it has been argued that those have a negligible influence and can thus be ignored \citep{Fan2006, Becker2015}. 

\begin{figure}[!t]
\centering
\includegraphics[width=.47\textwidth]{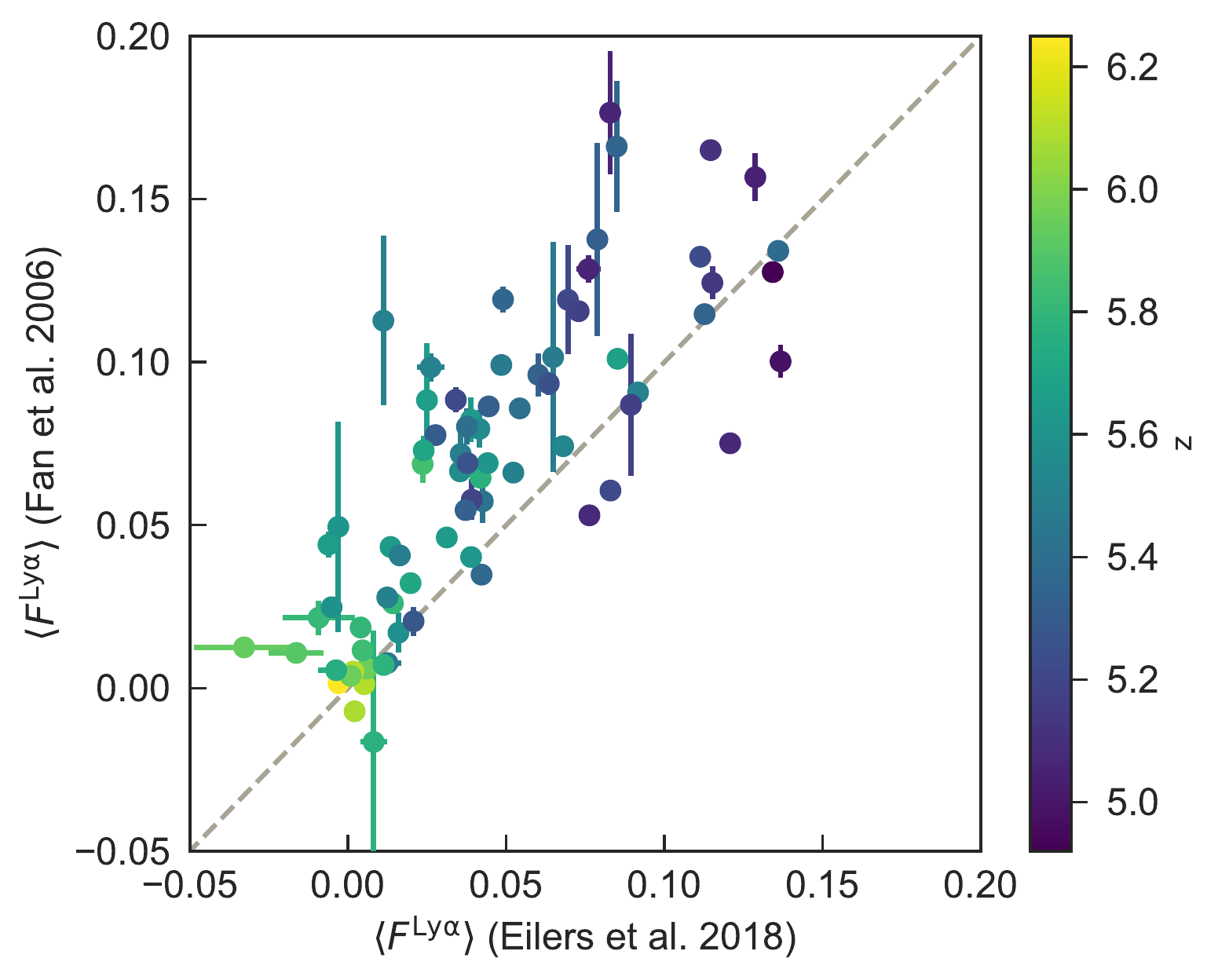}
\caption{Mean flux measurements within the \lya forest of \citet{Fan2006} plotted vs. our measurements within the same redshift bins of $\Delta z = 0.15$ for the $16$ quasar sightlines that are part of both data sets.  \label{fig:sys_comp_fan}} 
\end{figure}

All these differences contribute to the systematic error of the opacity measurements. We attempted to assess these systematic uncertainties by comparing the results from our analysis to \citet{Fan2006} along the sight lines that are part of both data sets. To this end we measure the mean flux in the same redshift bins as \citet{Fan2006} with a fixed bin size of $\Delta z = 0.15$, and compare our measurements to \citet{Fan2006} in Fig.~\ref{fig:sys_comp_fan}. We observe a large scatter in the distribution and a systematic offset towards lower mean flux values in our measurements, much larger than the formal measurement uncertainties. The negative offset is strongest at lower redshifts with higher mean flux values, i.e. lower optical depths. 

We estimate the systematic error arising due to different observations, different data reduction pipelines and different analyses by the median of the distribution of measured flux differences $\Delta \langle F\rangle= \langle F_{\rm Eilers\,et\,al.\,2018}\rangle - \langle F_{\rm Fan\,et\,al.\,2006}\rangle$. We find a median systematic error of
\begin{align*}
\sigma_{\Delta\langle F\rangle} \approx -0.023,    
\end{align*}
with a large scatter of $\approx 0.026$ determined from the mean of the $16$th and $84$ percentile of the distribution. A detailed investigation of a few of the largest outliers in this distribution suggests that differences in the spectra itself, due to the different instruments with which they were observed and potentially due to differences in the data reduction, cause the largest discrepancies. 

Recently, a similar analysis of the
\lya optical depths measured from a comparable
quasar sample at $z_{\rm em} > 5.7$ was presented by 
\citet{Bosman2018}. Of the $62$ sight lines they analyzed, 
$22$ satisfy our quality criteria, namely that they
are non-BAL quasars with a $\rm S/N > 7$\footnote{Note that \citet{Bosman2018} quote a
  $\rm S/N$ ratio per $60\,\rm km\,s^{-1}$ pixel. Our $\rm S/N$ quoted in Tab.~\ref{tab:overview_data} is calculated per $10\,\rm km\,s^{-1}$ pixel so for a direct comparison we have to correct our quoted $\rm S/N$ ratios by $\rm S/N_{60\,\rm km\,s^{-1}} = S/N_{10\,\rm km\,s^{-1}} \cdot \sqrt{6}$, i.e. the threshold we apply for including spectra in our analysis is $\rm S/N_{60\,\rm km\,s^{-1}} > 17.1$. }. 
Out of these, $17$ objects overlap with our sample. 
Although the \citet{Bosman2018} sample is comparable to ours in size and partially overlapping, their methods differ in a variety of important aspects from ours.  
As in \citet{Fan2006} different data reduction pipelines have been used to reduce the data, the quasar continuum estimation methods differ, and while both our study and \citet{Bosman2018} mask the proximity zone regions, we adapt the excluded region dependent on the actual measured proximity zone size $R_p$ (see \S~\ref{sec:measurements}), whereas their analysis excludes a fixed spectral range until $\lambda_{\rm rest} = 1178\,${\AA}, which corresponds to $\Delta R_p = 13.3\,$pMpc at $z=6$. Finally, we have masked strong absorbers and account for small zero-level offsets, whereas they do not. 

\subsection{Comparison of the Cumulative Distribution Functions}\label{sec:comp_cdf}

\begin{figure*}[!t]
\centering
\includegraphics[width=\textwidth]{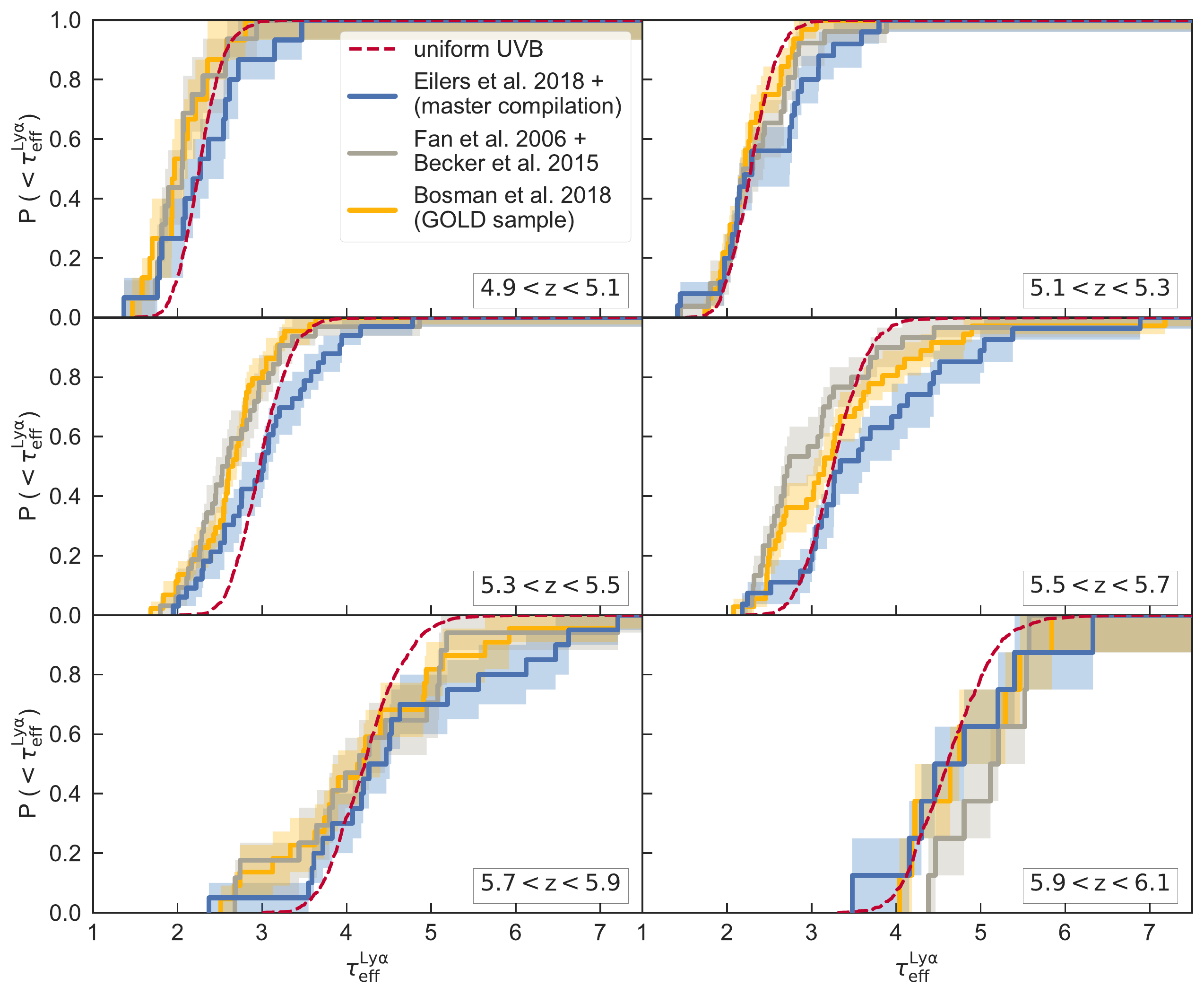}
\caption{Cumulative distribution functions (CDF) of $\tau_{\rm eff}^{\rm Ly\alpha}$ in different redshift bins. The blue curves show the master compilation of our new measurements with $\rm ULAS\,J1048+0600$ from \citet{Becker2015}, whereas the grey and yellow curves show the CDF from previous measurements \citep{Fan2006, Becker2015} as well as the  new compilation by \citet{Bosman2018}, respectively. The shaded regions show the $1\sigma$-uncertainties determined via boostrapping. The red dashed curves show CDFs from hydrodynamical simulations assuming a uniform UVB. Note that the simulations have been re-scaled to match the mean \taueff in the respective bin by applying eqn.~\ref{eq:rescale} and the procedure described in \S~\ref{sec:sims_lya}. \label{fig:cdf_alpha}} 
\end{figure*}

In Fig.~\ref{fig:cdf_alpha} we compare the CDF from our measurements shown in blue to the CDF from previous studies by \citet{Fan2006} and \citet{Becker2015}, which are shown as the grey curves, and by \citet{Bosman2018} shown in yellow, in different redshift bins centered around $5.0 \leq z\leq 6.0$. We show the so-called GOLD sample from \citet{Bosman2018} including $33$ quasar spectra for which they applied a data quality cut of $\rm S/N > 11.2$ per $60\,\rm km\,s^{-1}$ pixel to their sample, which would imply a quality cut on our sample of $\rm S/N > 4.6$ per $10\,\rm km\,s^{-1}$ pixel. 

While previous work noticed an increased scatter in the opacity measurements only at $z\gtrsim 5.5$, we also see evidence for increased scatter at $5.0 < z < 5.5$. We see systematic differences towards higher optical depths in our work compared to others, most strikingly in the $5.3 < z < 5.7$ bins whose excess fluctuations have been the focus of several works. However, in most redshift bins the measurements agree within $1\sigma$-uncertainties (shown as the shaded regions in Fig.~\ref{fig:cdf_alpha}) which we determined via bootstrap resampling, the only exception being the redshift bin at $z=5.4$ and $z=5.6$, where our results are slightly more discrepant with previous studies. 

A discrepancy between the \citet{Fan2006} and \citet{Becker2015} measurements in this bin was previously noted by \citet{Chardin2017}, and in particular it seems that our (higher) \taueff measurements are more consistent with the data from \citet{Becker2015} than those from \citet{Fan2006}.  

\section{Simulations of the IGM}\label{sec:sims}

We would like to compare our measurements of the IGM opacity to expectations from simulations. 
For this purpose we use a hydrodynamical simulation which we briefly describe in section \S~\ref{sec:nyx}. In this simulation we use a uniform UVB radiation field and thus this simulations provides a good approximation for opacity fluctuations in the IGM long after the epoch of reionization, when we expect to have a uniform UVB. We use two more sophisticated models to compare our measurements with more realistic conditions in the post-reionization IGM. To this end, we use two semi-numerical models with fluctuating UVB and temperature field, which we describe in \S~\ref{sec:fluct_models}.  
In \S~\ref{sec:sims_lya} we explain how we compute the \lya optical depth from the skewers through the simulation box. 

\begin{figure}[!t]
\centering
\includegraphics[width=.47\textwidth]{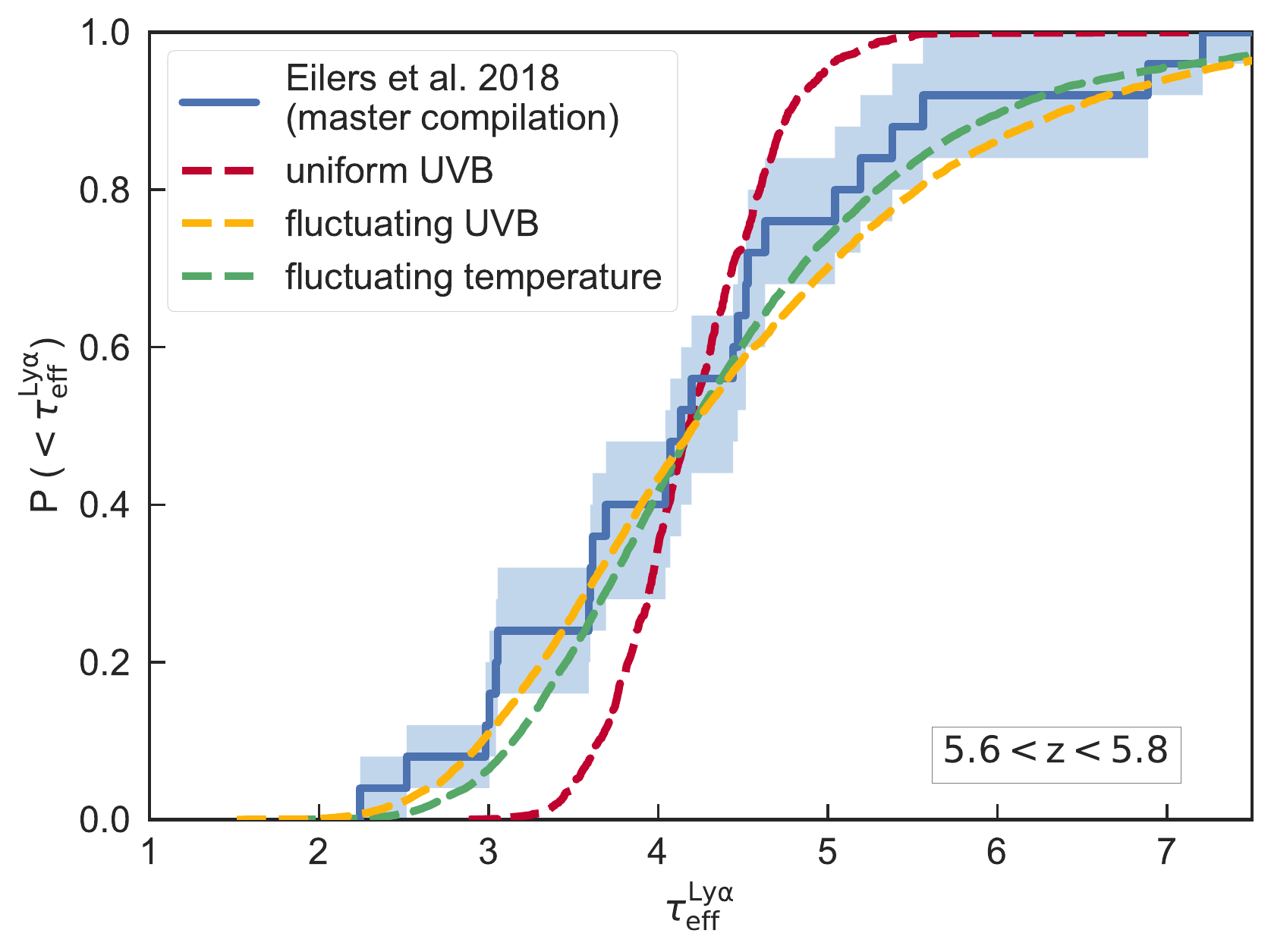}
\caption{The observed CDF at $z=5.7$ (blue curve) compared to a hydrodynamical simulation with a uniform UVB model (red curve) and a semi-numerical simulation with either a fluctuating UVB (yellow curve) or fluctuating post-reionization temperature (green curve). The blue shaded regions shows the $1\sigma$-uncertaintiy on the CDF determined via boostrapping. \label{fig:cdf_fred}} 
\end{figure}

\subsection{Nyx Hydrodynamical Simulation}\label{sec:nyx}

In this work we employ a \texttt{Nyx} hydrodynamical simulation \citep{Almgren2013} 100 \cmpch\ on a side with 4096$^3$ dark matter particles and gas elements on a uniform Eulerian grid, designed for precision studies of the \lya forest \citep{Lukic2015}. 
We extracted 1000 random skewers
 of density, temperature, and velocity along the directions of the grid axes from simulation outputs at $z=$3.0, 3.5, 4.0, 4.5, 5.0, 5.5, 6.0, and 6.5. We then computed the resulting \lya forest spectra in 50 \cmpch\ bins consistent with the scale of the IGM opacity measurements presented here. For redshift bins in between the simulation outputs, we take the closest output and re-scale the density field by $(1+z)^3$ accordingly. 

The simulation adopted the uniform UVB model of \citet{FaucherGiguere2009}, resulting in a ``vanilla" IGM model which (uniformly) reionized at early times ($z_{\rm reion}>10$). Thus any deviations of the distribution of IGM opacity observations from that predicted in the simulation likely represent spatial inhomogeneities in the UVB \citep{DaviesFurlanetto2016} or thermal state related to a more recent epoch of reionization \citep{Daloisio2015}. 

\subsection{Semi-Numerical Models with Fluctuating UVB and Temperature Fields}\label{sec:fluct_models}

We also compare our observations at $z=5.7$ to the semi-numerical fluctuating UVB and fluctuating IGM temperature models from \citet{Davies2017} which we describe briefly below.

The \citet{Davies2017} semi-numerical simulation consists of a cosmological volume, 780 cMpc on a side, with a 2048$^3$ density field computed via the Zel'dovich approximation \citep{Zeldovich1970} and dark matter halos ($M_{\rm halo}\geq2\times10^9$ M$_\odot$) populated using the excursion set formalism as in \citet{MesingerFurlanetto2007}. Ionizing luminosities were assigned to halos by first abundance matching to the (non-ionizing) UV luminosity function \citep{Bouwens2015} and then allowing the ratio of ionizing to non-ionizing luminosities to vary as a free parameter. 

UVB fluctuations in this volume were computed on a 156$^3$ grid following the method of \citet{DaviesFurlanetto2016} with a spatially-varying mean free path of ionizing photons. To construct a fluctuating IGM temperature field, the reionization redshifts of each density cell in the semi-numerical simulation were computed with \texttt{21cmFAST} \citep{Mesinger11} and the subsequent cooling from an initial post-reionization temperature of $30,000$ K was computed via numerical integration of the IGM thermal evolution (as in \citealt{UptonSanderbeck2016}). Finally, Ly$\alpha$ forest sightlines were then computed using the fluctuating Gunn-Peterson approximation \citep{Weinberg1997} applied to the Zel'dovich approximation density field, with a normalization factor applied to the optical depth of each pixel to account for the approximate nature of the method.

\subsection{Calculating the \lya Optical Depth from Simulated Skewers}\label{sec:sims_lya}

We then extract $1000$ skewers through the various simulation boxes to compute the optical depths and compare the results to our measurements. 
Because the exact strength of the UVB radiation $\Gamma_{\rm UVB}$ is unknown, we have to re-scale the optical depth in each skewer at each pixel $i$, i.e. $\tau_{i}^{\rm Ly\alpha, unscaled}$, to match the mean optical depth corresponding to the observed mean flux value $\langle F^{\rm obs}\rangle$ of our measurements, which in turn depends on the exact value of $\Gamma_{\rm UVB}$. Hence, at each redshift we determine a scaling factor $A_0$ that solves the following equation:
\begin{align}
\langle \exp\left[-\tau_{i}^{\rm Ly\alpha}\right]\rangle &= \langle \exp\left[-A_0 \times \tau_{i}^{\rm Ly\alpha, unscaled}\right]\rangle\nonumber\\
& = \langle F^{\rm obs}\rangle\label{eq:rescale}
\end{align}
We then average the re-scaled flux at each pixel $\langle \exp\left[-\tau_{i}^{\rm Ly\alpha}\right]\rangle$ within each skewer of size $50$\,\cmpch, 
and determine the $68\%$ and $95\%$ of the distribution of $\tau_{\rm eff}^{\rm Ly\alpha}$.
 This gives an estimate of the expected scatter within the \lya optical depth measurements given a uniform UVB and IGM thermal state, that is, arising from density fluctuations alone.

\section{Implications for the Epoch of Reionization}\label{sec:discussion}

\begin{figure}[!t]
\centering
\includegraphics[width=.47\textwidth]{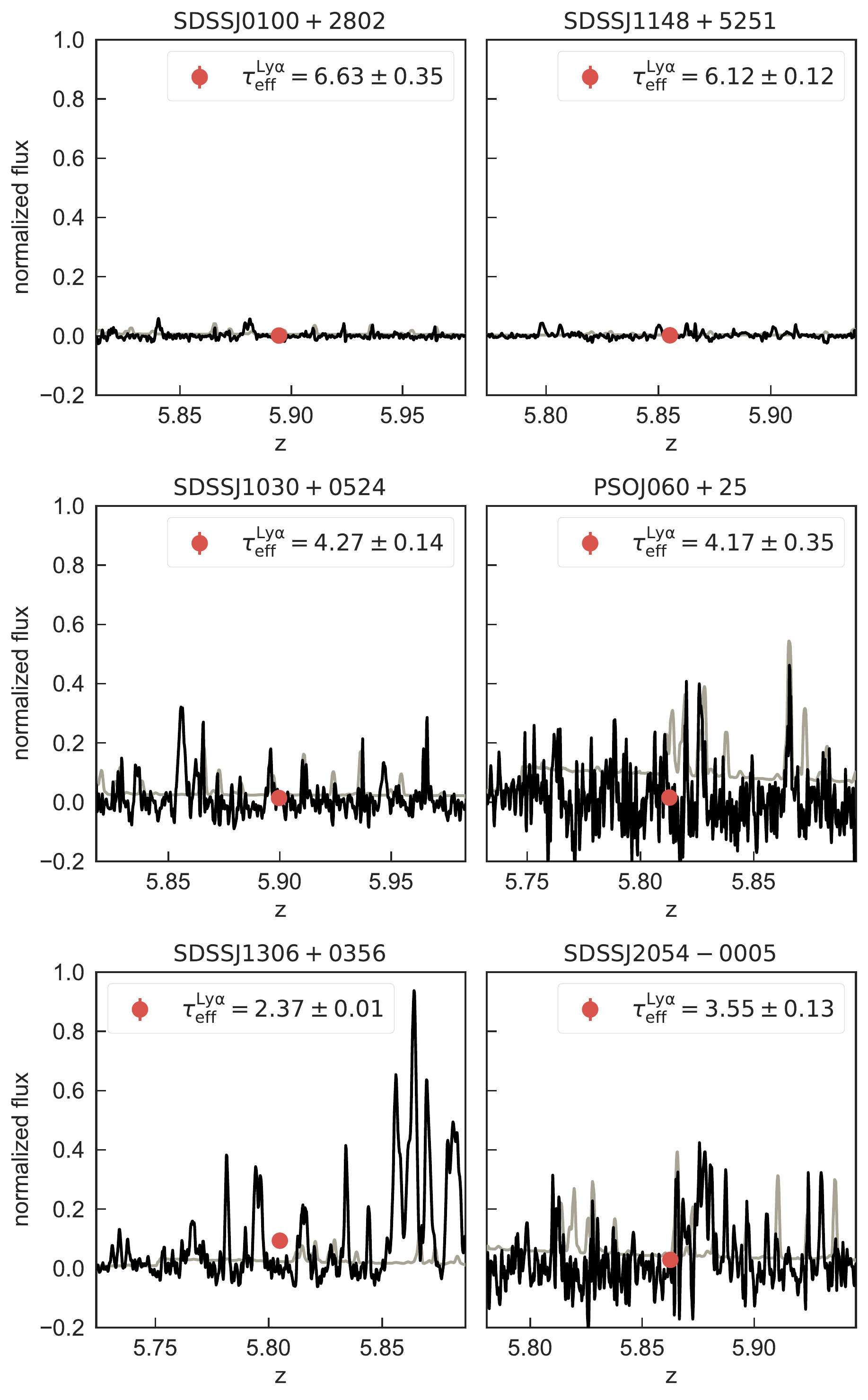}
\caption{Selected spectral bins of $50\,$\cmpch at a similar redshift, for which we measure $\tau_{\rm eff}^{\rm Ly\alpha}$ within the \lya forest, along six different quasar sight lines that demonstrate the observed scatter. The sight lines in the top panels exhibit very low transmitted flux, i.e. very high optical depths, the middle panels show example spectral bins for a medium level of transmitted flux, whereas the sight lines in the bottom panels show abundant transmitted flux, i.e. very low optical depths. All spectral bins shown are at comparable redshifts. The grey curve in each panel shows the respective noise vectors and the red data points show the measurements of $\langle F^{\rm obs}\rangle$ (the errorbar is smaller than the symbol and thus not visible) with the corresponding optical depth measurements in the legend. \label{fig:chunk_alpha}} 
\end{figure}

In order to assess the implications of our opacity measurements for the epoch of reionization, we compare our measurements to the opacity distributions from the \texttt{Nyx} hydrodynamical simulation (\S~\ref{sec:nyx}). The \texttt{Nyx} simulation was computed with a uniform UVB, which we would expect long after the epoch of reionization or when assuming no signatures of an inhomogeneous reionization process. The evolution of the optical depth distributions from these simulations, with the mean fluxes matched to our measurements, are shown as the grey region in Fig.~\ref{fig:tau_alpha}. 
The width of the light ($68$th percentile) and medium ($95$th percentile) grey regions corresponds to the scatter in $\tau_{\rm eff}$ expected due to fluctuations in the underlying density field alone. 
The dark grey regions indicate the additional scatter expected from $\sim 20\%$ uncertainties in the quasar continuum estimation. These have been calculated by dividing the mean flux in each spectral bin by $(1 + \Delta C)$, where the continuum uncertainty $\Delta C$ was drawn randomly from a normal distribution with $\sigma = 0.2$ and $\mu = 0$, corresponding to $\sim 20\%$ uncertainties in the continuum estimate. As expected continuum uncertainties matter only very little at higher redshifts, when the mean transmitted flux is low and fluctuations between different quasar sightlines dominate, and the scatter at high redshift can thus not be explained by continuum uncertainties. 

The measurements show a steep rise in \taueff for $z\gtrsim 5$ and an increased scatter in the distribution of measurements. At lower redshifts ($z\lesssim 5$) the scatter in the observed \taueff decreases rapidly and becomes consistent with the expectations from density fluctuations alone. 
It is evident, however, that at high redshifts the scatter in the optical depths measurements significantly exceeds the scatter expected from density fluctuations alone, i.e. the scatter represented by our hydrodynamical simulation with uniform UVB.
The tiny rare flux spikes observed in the \lya forests of $\rm SDSS~J0100+2802$, and $\rm SDSS~J1148+5251$ at redshifts $5.8 \lesssim z \lesssim 5.9$ are in strong contrast to the abundant transmitted flux along the sight lines towards $\rm SDSS~J1306+0356$ or $\rm SDSS~J2054-0005$ at similar redshifts, for example. 
We show the respective spectral regions exhibiting very high (upper panels), average (middle panels) and very low (lower panels) optical depths at similar redshifts of the aforementioned sight lines in Fig.~\ref{fig:chunk_alpha}. 

The discrepancy between our measured opacity distribution and the expectation from simulations with a uniform UVB becomes even more evident in Fig.~\ref{fig:cdf_alpha}, where we show the cumulative distributions of $\tau_{\rm eff}^{\rm Ly\alpha}$ from our master compilation in different redshift bins. The CDFs of \taueff from our 
hydrodynamical simulation $+$ uniform UVB are shown as the dashed red curves, where we have rescaled the pixel optical depths (\S~\ref{sec:sims_lya}) to match the mean \taueff in each redshift bin. 
This model with a uniform UVB is clearly not a good match to the observations. While they are more consistent with the measurements at lower redshift ($z\sim 5$), there are large discrepancies at higher redshifts ($z \gtrsim 5.6$) between the simulated and the observed CDF, a point previously noted by \citet{Fan2006} and \citet{Becker2015}. 
While it may seem that the model provides a better fit in the highest redshift bin at $z\sim 6.0$, this apparent agreement is
misleading, and arises due to the fact that we show limits on the optical depth in the same way as measurements, and the bin at $z\sim6$ contains several limits.

\subsection{Comparison to Patchy Reionization Models}

Multiple scenarios have been proposed to explain the increased scatter in the optical depth relative to the fluctuations one would expect from the density field of the IGM alone. One possible explanation is that the UVB is strongly fluctuating, either due to coupled variations in the mean free path of ionizing photons \citep{DaviesFurlanetto2016, Daloisio2018} or a rare source population, such as quasars \citep{Chardin2015, Chardin2017}. Another possibility is that the thermal state of the IGM is highly inhomogeneous \citep{LidzMalloy2014, Daloisio2015}. Such fluctuations can arise as a result of an extended and patchy reionization process, where different regions in the universe were reionized (and simultaneously photoheated) at different redshifts $z_{\rm reion}$. The regions that reionized earlier would have had time to cool down while regions that were reionized at a later time would still be relatively hot.

In Fig.~\ref{fig:cdf_fred} we compare our measurements at $5.6 < z < 5.8$ to the semi-numerical models with a fluctuating UVB and fluctuating temperature field (see \S~\ref{sec:fluct_models}), where we have re-scaled the opacities in the \lya forest skewers from that work to match the mean \taueff we have measured in this bin.
Note that while previously the CDF of optical depths at this redshift bin containing the strong outliers in the opacity measurements in the GP trough along the sightline of $\rm ULAS\,J0148+0600$ \citep{Becker2015}, was very challenging to reproduce in simulations \citep{Daloisio2015, DaviesFurlanetto2016, Davies2017, Chardin2017, Keating2017}, these outliers are now easier to explain because the mean \taueff in our measurements is substantially higher than the \citet{Fan2006}+\citet{Becker2015} compilation, and hence these data points do not represent such strong deviations from the mean of the distribution anymore. 
This first comparison of our measurements to the two semi-numerical models with a fluctuating UVB and a fluctuating thermal state of the IGM shows that both models can reproduce the observations. A more detailed comparison to these models will be part of future work.

\section{Summary}\label{sec:summary}

We present a new data set of $34$ quasar spectra at $5.77\leq z_{\rm em} \leq 6.54$ that we make publicly available via the \texttt{igmspec} database. The spectra have all been observed with ESI on the Keck telescopes, and exposures from different observing runs have been co-added, resulting in a very rich and homogeneous data set, with a total of $\sim 180$ hours of telescope time. 

For a subsample of $23$ quasar spectra, that do not show broad absorption line features and have good quality data (i.e. $\rm S/N > 7$), we measure the IGM opacity by means of the effective optical depth of the \lya 
forest in bins of $50$~\cmpch covering a redshift range of $4.0\lesssim z\lesssim 6.5$. 
Our results are in qualitative agreement with previous studies \citep{Fan2006, Becker2015, Bosman2018}, showing a steep rise in opacity and increased scatter within the measurements at high redshift. However, while previous work noticed an increased scatter at $z\gtrsim 5.5$, we also see evidence for increased scatter at $5.0 < z < 5.5$. A detailed comparison in the optical depth in several redshift bins, shows systematic differences towards higher optical depths in our work compared to others, most strikingly at $5.3<z<5.7$. The discrepancies, however, between our measurements and previous work are mostly within the $\sim 1\sigma$ uncertainties, which we determined via bootstrap resampling. 

Our work improves upon previous studies in several aspects. We considered possible contamination due to intervening low-ion metal absorption systems such as DLAs that have previously been ignored and carefully masked all regions in the \lya 
forest that are affected by these high \ion{H}{1} column density absorption systems. 
 We also corrected for small offsets in the zero-level of the quasar spectra, introduced presumably by improper sky subtraction of a few individual exposures. 
Finally and most importantly, we considered a 
very homogeneously reduced data sample which minimizes systematic effects due to the use of different telescopes and detectors, or data reduction pipelines. 
We present a master compilation sample including mainly our newly analyzed sample with the exception of the sightline of $\rm ULAS\,J0148+0600$ taken from \citet{Becker2015}, who has a larger sensitivity in the prominent GP trough due to the higher signal-to-noise ratio spectrum observed with VLT/X-Shooter. 

We compare our measurements to a large-volume hydrodynamical simulation with a uniform UVB. As noted previously by \citet{Fan2006} and \citet{Becker2015}, we find
that the spread in observed \taueff cannot be explained by fluctuations of the underlying density field alone, 
and thus our results support an inhomogeneous reionization scenario. Whether temperature fluctuations in the IGM, a fluctuating UVB or a combination of both can best explain this increased scatter in opacity, remains an open question. 
A preliminary comparison of our measurements to semi-numerical simulations of UVB and IGM temperature fluctuations shows good agreement for both scenarios.

This work presents a crucial ingredient in constraining the end-stages of the epoch of reionization at $5.0 \lesssim z \lesssim 6.0$, when the physical conditions of the post-reionization IGM can be directly measured via absorption spectroscopy of high redshift quasars. The past several years have seen an impressive fivefold increase in the number of $z > 6$ quasars from deep wide-field optical and infrared  surveys, which are enabling precise measurements of the \lya forest absorption at $5 < z < 6.5$ \citep{Becker2015, Gnedin2017, Davies2017}. The requirement that reionization models reproduce these high-precision measurements provides an important low redshift anchor point which all models must reproduce, and can dramatically narrow the family of viable reionization models. Statistical anaylses of the \lya forest, such as measurements of the power spectrum \citep{Onorbe2017, Daloisio2018} or the PDF of the IGM opacity \citep{DaviesHennawiEilers2017} set further constraints on the reionization process, allowing us to develop accurate models about the early evolutionary stages of our universe.

\acknowledgments

We would like to thank the anonymous referee for insightful and constructive feedback. We thank Jos\'e O\~{n}orbe for helpful discussions regarding the hydrodynamical simulations, Zarija Luki\'c for providing the \texttt{Nyx} simulation, and Vikram Khaire for valuable feedback. Furthermore, we appreciate the help of Xiaohui Fan, making his quasar spectra available to us for a data comparison. 

The data presented in this paper were obtained at the W.M. Keck
Observatory, which is operated as a scientific partnership among the
California Institute of Technology, the University of California and
the National Aeronautics and Space Administration. The Observatory was
made possible by the generous financial support of the W.M. Keck
Foundation.

This research has made use of the Keck Observatory Archive (KOA), which is operated by the W. M. Keck Observatory and the NASA Exoplanet Science Institute (NExScI), under contract with the National Aeronautics and Space Administration.

The authors wish to recognize and acknowledge the very significant cultural role and reverence that the summit of Mauna Kea has always had within the indigenous Hawaiian community.  We are most fortunate to have the opportunity to conduct observations from this mountain.

\bibliography{literatur_hz}

\software{ESIRedux (http://www2.keck.hawaii.edu/inst/esi/ESIRedux/), XIDL (http://www.ucolick.org/~xavier/IDL/), igmspec (http://specdb.readthedocs.io/en/latest/igmspec.html), Nyx \citep{Almgren2013}, emcee \citep{emcee}, numpy \citep{numpy}, scipy \citep{scipy}, matplotlib \citep{matplotlib}, astropy \citep{astropy}, seaborn \citep{seaborn}}

\appendix

\section{Catalog of the Data Release}\label{sec:catalog}
In Tab.~\ref{tab:catalog} we present the catalog with the properties of our data set that will be available together with the final co-added spectra, their noise vectors and their continuum estimates via the \texttt{igmspec} database\footnote{\url{http://specdb.readthedocs.io/en/latest/igmspec.html}}. 
\texttt{igmspec} is a database of publicly available ultraviolet, optical, and near-infrared spectra that probe the IGM. It provides $\sim 500,000$ spectra from various datasets including the Sloan Digital Sky Survey (SDSS), 2dF QSO Redshift Survey (2QZ), and data from the Hubble Space Telescope (HST), the Keck Telescopes, the Very Large Telescope (VLT), and more. The database is part of the \textit{specdb} repository\footnote{\url{http://specdb.readthedocs.io/en/latest/}} within the \textit{specdb} Github organization\footnote{\url{https://github.com/specdb/specdb}}, which provides software developed in Python for accessing and interacting with the quasar spectra.

\startlongtable
\begin{deluxetable*}{lllLLcccc}
\tablewidth{\textwidth}
\tablecaption{Catalog for the \texttt{igmspec} data base. \label{tab:catalog}}
\tablehead{\colhead{object} & \colhead{RA [deg]} & \colhead{DEC [deg]} & \dcolhead{z_{\rm em}} & \dcolhead{M_{\rm 1450}} & \colhead{S/N} & \dcolhead{R [\Delta\lambda/\lambda]} &\colhead{telescope} &\colhead{instrument}}
\startdata
SDSS\,J0002+2550 & 0.6641 & 25.843 & 5.82 & -27.31 & 57 & 5400 & Keck II & ESI \\
SDSS\,J0005-0006 & 1.4681 & -0.1155 & 5.844 & -25.73 & 13 & 4000 & Keck II & ESI \\
CFHQS\,J0050+3445 & 13.7621 & 34.756 & 6.253 & -26.7 & 18 & 4000 & Keck II & ESI \\
SDSS\,J0100+2802 & 15.0542 & 28.0405 & 6.3258 & -29.14 & 35 & 4000 & Keck II & ESI \\
ULAS\,J0148+0600 & 27.1568 & 6.0056 & 5.98 & -27.39 & 20 & 4000 & Keck II & ESI \\
ULAS\,J0203+0012 & 30.8849 & 0.2081 & 5.72 & -26.26 & 5 & 4000 & Keck II & ESI \\
CFHQS\,J0210-0456 & 32.555 & -4.9391 & 6.4323 & -24.53 & 1 & 5400 & Keck II & ESI \\
PSO\,J0226+0302 & 36.5078 & 3.0498 & 6.5412 & -27.33 & 9 & 4000 & Keck II & ESI \\
CFHQS\,J0227-0605 & 36.9304 & -6.0917 & 6.2 & -25.28 & 3 & 4000 & Keck II & ESI \\
SDSS\,J0303-0019 & 45.8808 & -0.3202 & 6.078 & -25.56 & 2 & 4000 & Keck II & ESI \\
SDSS\,J0353+0104 & 58.4572 & 1.068 & 6.072 & -26.43 & 10 & 4000 & Keck II & ESI \\
PSO\,J0402+2451 & 60.5529 & 24.8568 & 6.18 & -26.95 & 11 & 4000 & Keck II & ESI \\
SDSS\,J0818+1722 & 124.6142 & 17.3811 & 6.02 & -27.52 & 5 & 4000 & Keck II & ESI \\
SDSS\,J0836+0054 & 129.1827 & 0.9148 & 5.81 & -27.75 & 108 & 4000 & Keck II & ESI \\
SDSS\,J0840+5624 & 130.1471 & 56.4056 & 5.8441 & -27.24 & 23 & 4000 & Keck II & ESI \\
SDSS\,J0842+1218 & 130.6226 & 12.314 & 6.069 & -26.91 & 8 & 4000 & Keck II & ESI \\
SDSS\,J0927+2001 & 141.8409 & 20.0232 & 5.7722 & -26.76 & 6 & 4000 & Keck II & ESI \\
SDSS\,J1030+0524 & 157.613 & 5.4153 & 6.309 & -26.99 & 20 & 4000 & Keck II & ESI \\
SDSS\,J1048+4637 & 162.1878 & 46.6218 & 6.2284 & -27.24 & 42 & 4000 & Keck II & ESI \\
SDSS\,J1137+3549 & 174.3239 & 35.8325 & 6.03 & -27.36 & 18 & 4000 & Keck II & ESI \\
SDSS\,J1148+5251 & 177.0694 & 52.864 & 6.4189 & -27.62 & 28 & 4000 & Keck II & ESI \\
SDSS\,J1250+3130 & 192.7164 & 31.5061 & 6.15 & -26.53 & 7 & 4000 & Keck II & ESI \\
SDSS\,J1306+0356 & 196.5345 & 3.9907 & 6.016 & -26.81 & 28 & 4000 & Keck II & ESI \\
ULAS\,J1319+0950 & 199.7971 & 9.8476 & 6.133 & -27.05 & 8 & 4000 & Keck II & ESI \\
SDSS\,J1335+3533 & 203.9617 & 35.5544 & 5.9012 & -26.67 & 5 & 4000 & Keck II & ESI \\
SDSS\,J1411+1217 & 212.797 & 12.2937 & 5.904 & -26.69 & 25 & 4000 & Keck II & ESI \\
SDSS\,J1602+4228 & 240.7249 & 42.4736 & 6.09 & -26.94 & 16 & 4000 & Keck II & ESI \\
SDSS\,J1623+3112 & 245.8825 & 31.2001 & 6.2572 & -26.55 & 9 & 4000 & Keck II & ESI \\
SDSS\,J1630+4012 & 247.6412 & 40.2027 & 6.065 & -26.19 & 11 & 4000 & Keck II & ESI \\
CFHQS\,J1641+3755 & 250.3405 & 37.9223 & 6.047 & -25.67 & 2 & 4000 & Keck II & ESI \\
SDSS\,J2054-0005 & 313.527 & -0.0874 & 6.0391 & -26.21 & 12 & 4000 & Keck II & ESI \\
CFHQS\,J2229+1457 & 337.2569 & 14.9525 & 6.1517 & -24.78 & 2 & 4000 & Keck II & ESI \\
SDSS\,J2315-0023 & 348.944 & -0.3995 & 6.117 & -25.66 & 12 & 4000 & Keck II & ESI \\
CFHQS\,J2329-0301 & 352.2845 & -3.033 & 6.417 & -25.25 & 2 & 4000 & Keck II & ESI \\
\enddata
\tablecomments{The columns show the object name, the coordinates $\rm RA$ and $\rm DEC$ of the quasar given in degrees, the emission redshift and 
the quasar's magnitude $M_{\rm 1450}$, the $\rm S/N$ of the data, the telescope and instrument with which the spectra are observed and the spectral resolution of the data. Note that in some cases we co-added data with different spectral resolution. }
\end{deluxetable*}

\section{Details of the Correction of Zero-Level Offsets in the Quasar Spectra}\label{sec:offset_details}

In order to correct for possible offsets in the zero-level of the quasar spectra, we examine the negative pixels in each spectral bin, which should appear to be a truncated Gaussian distribution, providing an estimate of the noise level in the spectra. To this end, we take all flux pixels $F$ below zero in each spectral bin of $50$~\cmpch, and calculate the cumulative distribution function (CDF), ignoring correlations between neighboring pixels. 

\begin{figure}[!t]
\centering
\includegraphics[width=.47\textwidth]{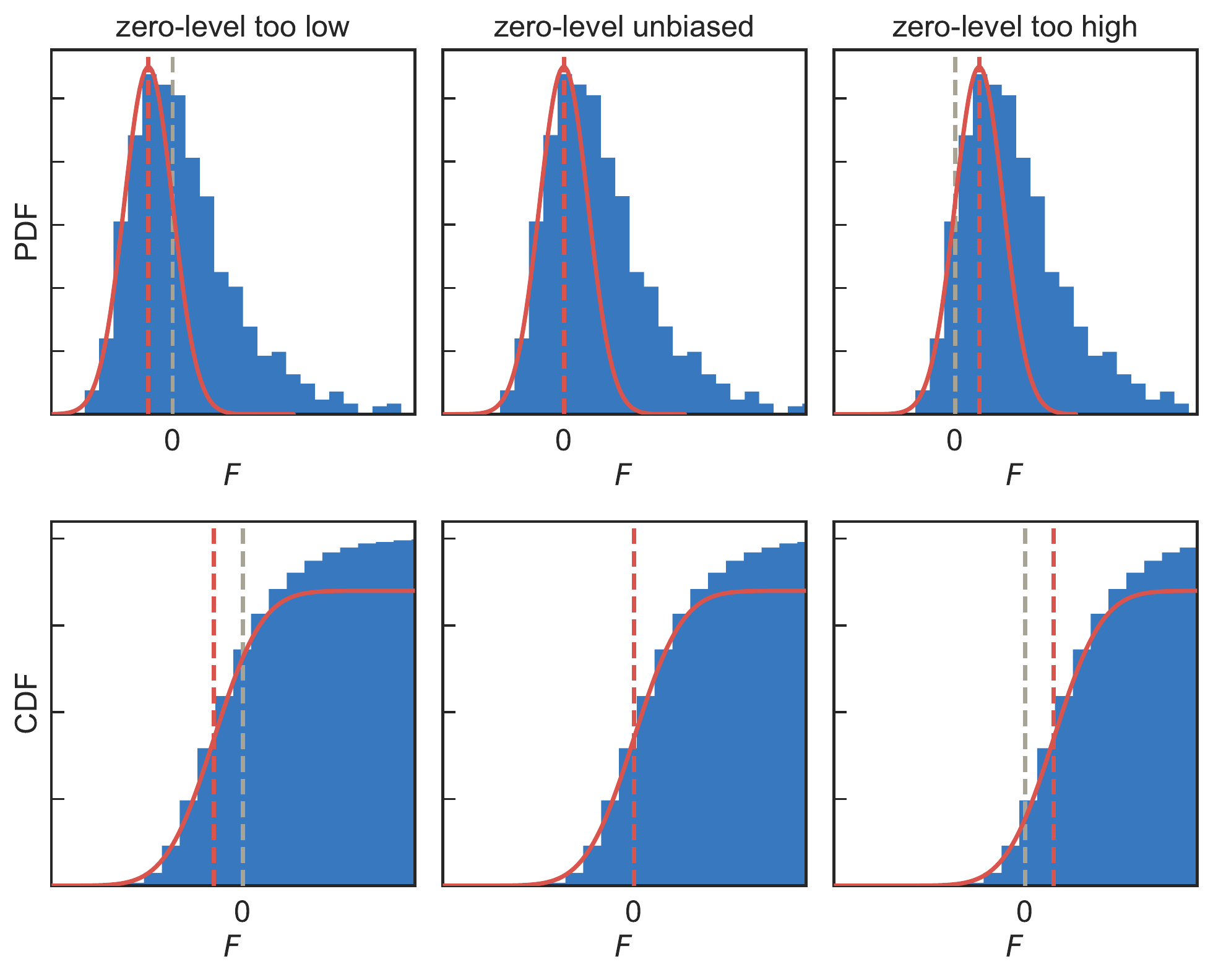}
\caption{Schematic depiction of the correction for offsets in the zero-level. The blue histograms show the PDF (top panels) and CDF (lower panels) of flux pixels from a toy model in a spectral bin. The red curves show a normal distribution of the expected noise level, with a mean indicated by the red dashed line. The grey dashed lines in the right and left panels show the current over- or underestimated zero-level, respectively.  \label{fig:zerolevel}} 
\end{figure}

In the case of no offset in the zero-level within a spectral bin, the
estimated mean of the CDF $\mu_{\rm CDF}$ should be equal to zero, as shown by the toy model example in the lower middle panel of
Fig.~\ref{fig:zerolevel}. The upper panels show the PDF of the same
respective toy model case. However, if the zero-level is slightly
under- or overestimated (left and right panels in
Fig.~\ref{fig:zerolevel}, respectively), the estimated $\mu_{\rm CDF}$
tracking the true zero-level will likewise be below or above zero.
It is clear that we can obtain a handle on these systematic offsets by examining the purely negative pixels and fitting a truncated CDF model. 

We use a Markov Chain Monte Carlo (MCMC) approach making use of the implementation of the affine-invariant ensemble sampler \texttt{emcee}\footnote{\url{http://dfm.io/emcee/current/}} \citep{emcee} to estimate the mean $\mu_{\rm CDF}$ of the best fitting model CDF
\begin{equation}
\text{CDF}_{\rm model}(F) = \frac{A_{\rm CDF}}{2}\,\left[1 + \text{erf}\left(\frac{F - \mu_{\rm CDF}}{\sqrt{2}\,\sigma_{\rm CDF}}\right)\right], 
\end{equation}
while marginalizing over the width of the distribution $\sigma_{\rm CDF}$ and the scaling factor $A_{\rm CDF}$. The likelihood function just maximizes the least squares between the CDF model and the measured CDF, i.e.
\begin{equation}
\ln \mathcal{L} = -0.5 \left(\rm CDF_{\rm model} - CDF_{\rm data}\right)^2. 
\end{equation}
We then take the median of the resulting posterior probability distribution function (PDF) as the best estimate for $\mu_{\rm CDF}$.

The three free parameters of the $\rm CDF_{\rm model}$, $\mu_{\rm CDF}$, $\sigma_{\rm CDF}$, and $A_{\rm CDF}$, are highly degenerate with each other because we are fitting only a small part of the CDF when taking solely flux pixels below zero into account, i.e. $F<0$. Thus we have to apply strict priors to break this degeneracy. The priors we chose are flat priors within the intervals $\mu_{\rm CDF}\in\left[-0.05,\,0.05\right]$, since we expect the total offset of the zero-level to be less than $\pm 5\%$, $\sigma_{\rm CDF}\in\left[0.75\,\sigma_{\rm eff},\,1.25\,\sigma_{\rm eff}\right]$, which takes into account the noise vector at each pixel $i$ of each quasar spectrum to estimate $\sigma_{\rm eff} = \sqrt{\sum_i\sigma_i^2/N}$, 
and $A_{\rm CDF}\in\left[0.45\,N_{F<0},\,0.55\,N_{F<0}\right]$.
The upper and lower boundaries for $A_{\rm CDF}$ result from the fact that the number of pixels with flux below zero, i.e. $N_{F<0}$, in an unbiased case should be exactly half of the pixels, i.e. $0.5\,N_{F<0}$. 
In the presence of the possible offsets in the zero-level of the spectra we allow $A_{\rm CDF}$ to deviate from the unbiased case. 

\begin{figure}
\centering
\includegraphics[width=.47\textwidth]{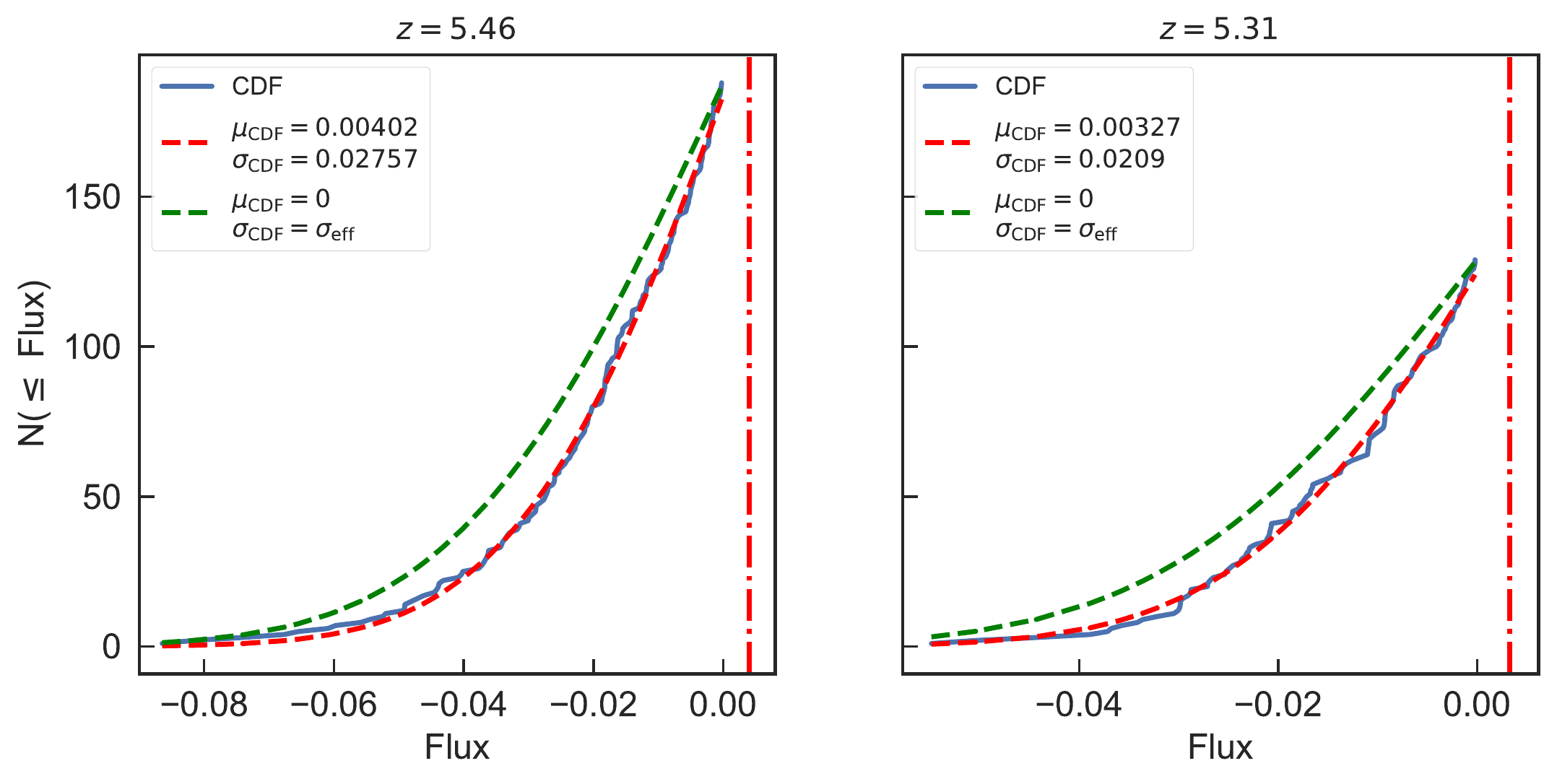}
\caption{Example CDF of two spectral bins along the sightline of $\rm SDSS~J0840+5624$. The blue curves show the measured CDF of each flux pixel, the red dotted curves show the best fitted CDF with the estimated values for mean and variance in the legend. The red dashed-dotted vertical line indicates the best fitted mean $\mu_{\rm CDF}$. The green dashed curves shows the ideal CDF, assuming no zero-level offset, i.e. $\mu_{\rm CDF} = 0$, and a correct noise model, i.e. $\sigma_{\rm CDF} = \sigma_{\rm eff}$.  \label{fig:cdf_ex}} 
\end{figure}

Fig.~\ref{fig:cdf_ex} shows two examples of the procedure. Both panels show the CDF for two spectral bins along the sightline of $\rm SDDS\,J0840+5624$. We show the actually measured CDF of all negative flux pixels (blue curves) and overplot the best fitted CDF (red dashed curves, with mean $\mu_{\rm CDF}$ indicated by the red dashed-dotted lines) and the ideal CDF with no zero-level offset, i.e. $\mu_{\rm CDF}=0$, and the variance given by the noise of the data, i.e. $\sigma_{\rm CDF} = \sigma_{\rm eff}$. We can see that we obtain small zero-level offsets of about $\Delta \langle F^{\rm Ly\alpha}\rangle \sim \mu_{\rm CDF}\approx 0.3-0.4\%$.

In the end, we offset all pixels $i$ within each spectral bin by the respective best estimate for $\mu_{\rm CDF}$, i.e. $F_{i,\,\rm new} = F_i - \mu_{\rm CDF}$, and calculate the mean flux and the opacity from the offseted pixels $F_{i,\,\rm new}$. 

\begin{figure}[!t]
\centering
\includegraphics[width=.47\textwidth]{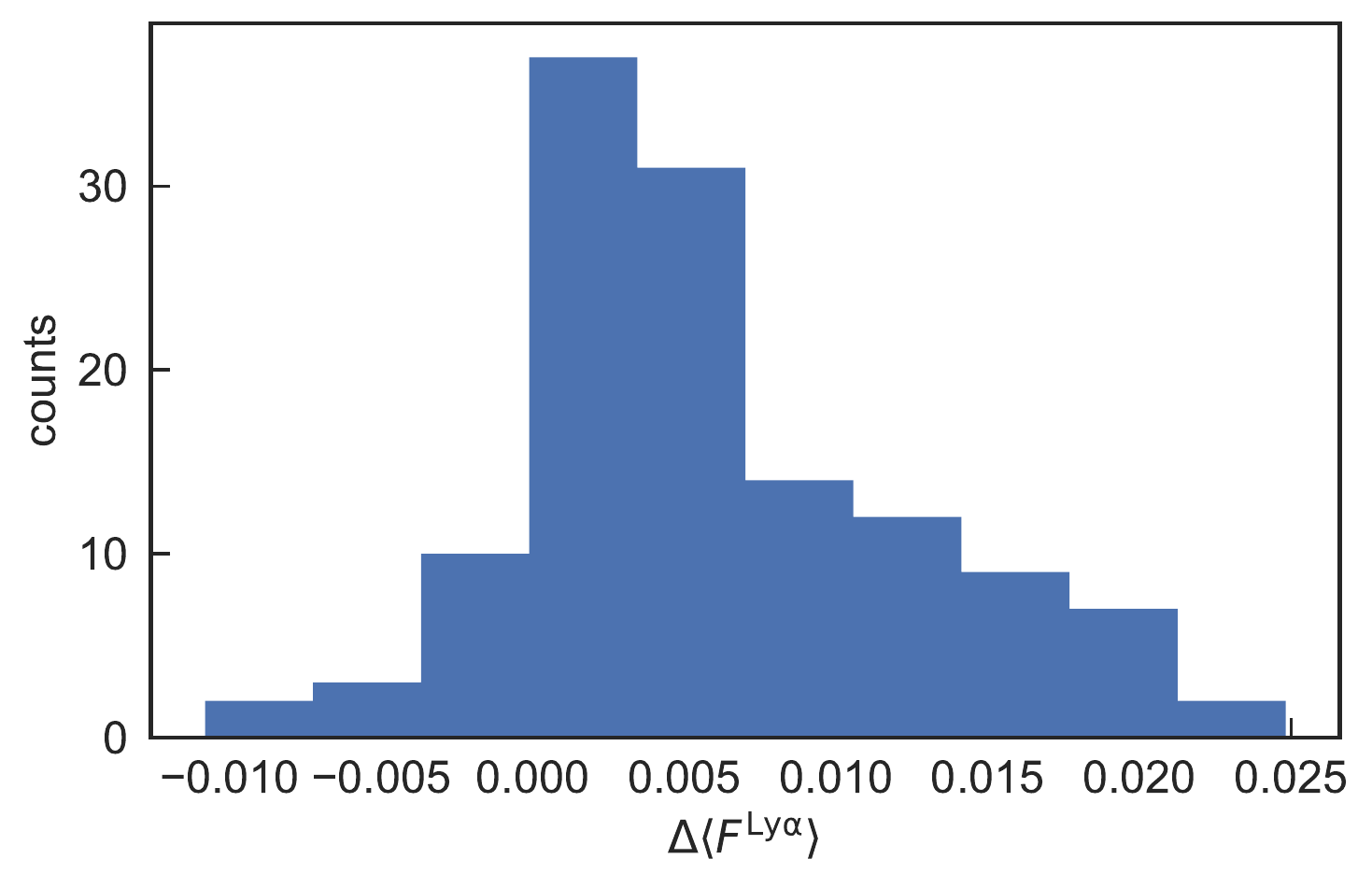}
\caption{Distribution of shifts in the mean flux estimate $\Delta \langle F^{\rm Ly\alpha}\rangle$ due to offsets in the zero-level of the quasar spectra.  \label{fig:sys_zero_level}} 
\end{figure}

Fig.~\ref{fig:sys_zero_level} shows the difference in mean flux estimates due to corrections in the zero-level. We estimate this systematic uncertainty in the mean flux measurements from calculating the $84$th percentile and $16$th percentile of this distribution and taking their average, which results in 
\begin{align*}
\sigma_{\langle F^{\rm Ly\alpha}\rangle} = 0.0067. 
\end{align*}

\section{All Quasar Continuum Estimates}\label{sec:all_cont_fits}

\begin{figure*}[ht!]
\centering
\includegraphics[width=0.98\textwidth]{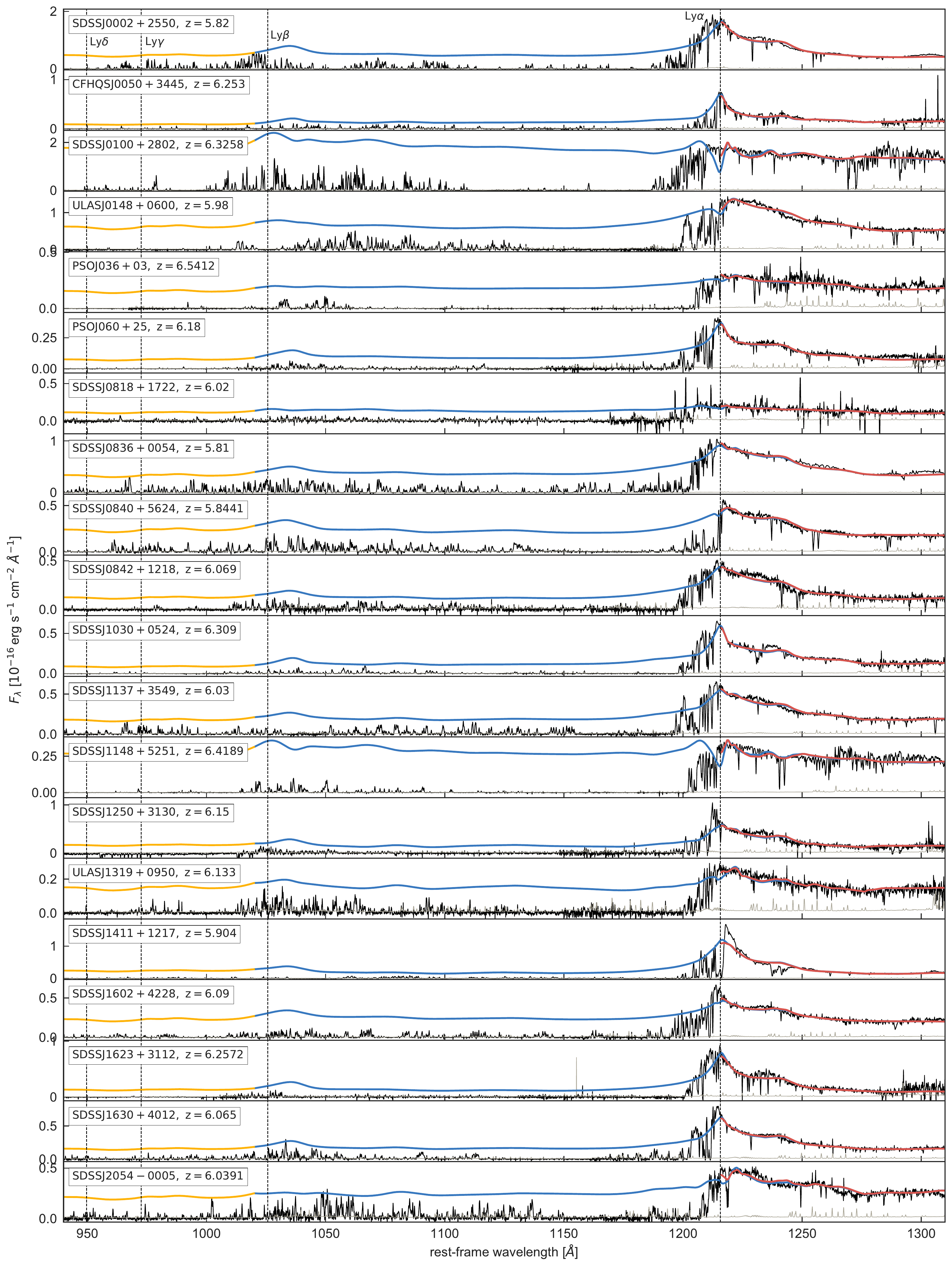}
\caption{Same as Fig.~\ref{fig:cont} for the remaining 20 quasar spectra in our data set used for the IGM opacity measurements. \label{fig:spectra_all_cont}} 
\end{figure*}

In Fig.~\ref{fig:spectra_all_cont} we show the estimates of the quasar continua for all objects, which we use for measurements of the IGM opacity that are not already shown in Fig.~\ref{fig:cont}. 

\section{Mean Flux Measurements within the \lya Forest}\label{sec:mean_flux}

Our measurements of the mean flux in the \lya 
forest are shown in Table~\ref{tab:lya_flux} 
for spectral bins of size $50$~\cmpch along all $23$ quasar sight lines in our data sample. 

\startlongtable
\begin{deluxetable}{lLLLRL}
\tabletypesize{\footnotesize}
\tablecaption{Mean flux measurements in the \lya forest. \label{tab:lya_flux}}
\tablehead{\colhead{object} & \dcolhead{z_{\rm start}} & \dcolhead{z_{\rm abs}} & \dcolhead{z_{\rm end}} & \dcolhead{\langle F\rangle} & \dcolhead{\sigma_{\langle F\rangle}}}
\startdata
SDSS\,J0002+2550 & 5.54 & 5.62 & 5.70 & 0.0472 & 0.0004 \\[-2 pt]
 & 5.39 & 5.47 & 5.54 & 0.0315 & 0.0005 \\
 & 5.25 & 5.32 & 5.39 & 0.0241 & 0.0004 \\
 & 5.10 & 5.17 & 5.25 & 0.1170 & 0.0005 \\
 & 4.97 & 5.04 & 5.10 & 0.1239 & 0.0006 \\
SDSS\,J0005-0006 & 5.60 & 5.68 & 5.76 & 0.0507 & 0.0016 \\[-2 pt]
 & 5.45 & 5.53 & 5.60 & 0.0384 & 0.0028 \\
 & 5.30 & 5.38 & 5.45 & 0.0484 & 0.0028 \\
 & 5.16 & 5.23 & 5.30 & 0.0460 & 0.0023 \\
 & 5.02 & 5.09 & 5.16 & 0.1128 & 0.0031 \\
 & 4.89 & 4.95 & 5.02 & 0.0732 & 0.0028 \\
CFHQS\,J0050+3445 & 5.96 & 6.05 & 6.14 & 0.0045 & 0.0018 \\[-2 pt]
 & 5.80 & 5.88 & 5.96 & 0.0217 & 0.0031 \\
 & 5.64 & 5.72 & 5.80 & 0.0269 & 0.0022 \\
 & 5.49 & 5.56 & 5.64 & 0.0570 & 0.0020 \\
 & 5.34 & 5.41 & 5.49 & 0.1336 & 0.0020 \\
 & 5.19 & 5.26 & 5.34 & 0.1009 & 0.0014 \\
SDSS\,J0100+2802 & 5.81 & 5.89 & 5.98 & 0.0013 & 0.0005 \\[-2 pt]
 & 5.65 & 5.73 & 5.81 & 0.0038 & 0.0003 \\
 & 5.50 & 5.57 & 5.65 & 0.0416 & 0.0003 \\
 & 5.35 & 5.42 & 5.50 & 0.0778 & 0.0003 \\
ULAS\,J0148+0600\tablenotemark{a} & 5.68 & 5.76 & 5.84 & 0.0056 & 0.0018 \\[-2 pt]
 & 5.53 & 5.60 & 5.68 & -0.0020 & 0.0016 \\
 & 5.38 & 5.45 & 5.53 & 0.0533 & 0.0018 \\
 & 5.23 & 5.30 & 5.38 & 0.0826 & 0.0014 \\
 & 5.09 & 5.16 & 5.23 & 0.1428 & 0.0014 \\
 & 4.95 & 5.02 & 5.09 & 0.1340 & 0.0018 \\
PSO\,J036+03 & 6.24 & 6.33 & 6.42 & -0.0018 & 0.0015 \\[-2 pt]
 & 6.06 & 6.15 & 6.24 & -0.0032 & 0.0011 \\
 & 5.89 & 5.98 & 6.06 & -0.0021 & 0.0009 \\
 & 5.73 & 5.81 & 5.89 & 0.0001 & 0.0008 \\
 & 5.57 & 5.65 & 5.73 & 0.0065 & 0.0006 \\
 & 5.42 & 5.50 & 5.57 & 0.0632 & 0.0012 \\
PSO\,J060+25 & 5.90 & 5.98 & 6.06 & 0.0308 & 0.0037 \\[-2 pt]
 & 5.73 & 5.81 & 5.90 & 0.0154 & 0.0054 \\
 & 5.57 & 5.65 & 5.73 & 0.0176 & 0.0020 \\
 & 5.42 & 5.50 & 5.57 & 0.0256 & 0.0023 \\
 & 5.28 & 5.35 & 5.42 & 0.0155 & 0.0017 \\
 & 5.13 & 5.20 & 5.28 & 0.0646 & 0.0016 \\
SDSS\,J0818+1722 & 5.56 & 5.64 & 5.72 & 0.0118 & 0.0034 \\[-2 pt]
 & 5.41 & 5.49 & 5.56 & 0.0194 & 0.0042 \\
 & 5.26 & 5.34 & 5.41 & 0.0462 & 0.0032 \\
 & 5.12 & 5.19 & 5.26 & 0.0457 & 0.0028 \\
 & 4.99 & 5.05 & 5.12 & 0.0768 & 0.0041 \\
SDSS\,J0836+0054 & 5.54 & 5.62 & 5.70 & 0.1061 & 0.0003 \\[-2 pt]
 & 5.39 & 5.46 & 5.54 & 0.0997 & 0.0004 \\
 & 5.24 & 5.31 & 5.39 & 0.0303 & 0.0003 \\
 & 5.10 & 5.17 & 5.24 & 0.1104 & 0.0003 \\
 & 4.97 & 5.03 & 5.10 & 0.1036 & 0.0004 \\
 & 4.83 & 4.90 & 4.97 & 0.1673 & 0.0003 \\
SDSS\,J0840+5624 & 5.39 & 5.46 & 5.54 & 0.0279 & 0.0012 \\[-2 pt]
 & 5.24 & 5.31 & 5.39 & 0.0778 & 0.0009 \\
 & 5.10 & 5.17 & 5.24 & 0.1010 & 0.0011 \\
 & 4.96 & 5.03 & 5.10 & 0.0937 & 0.0014 \\
 & 4.83 & 4.90 & 4.96 & 0.1487 & 0.0008 \\
SDSS\,J0842+1218 & 5.76 & 5.84 & 5.92 & -0.0177 & 0.0055 \\[-2 pt]
 & 5.60 & 5.68 & 5.76 & 0.0160 & 0.0037 \\
 & 5.44 & 5.52 & 5.60 & 0.0285 & 0.0052 \\
 & 5.30 & 5.37 & 5.44 & 0.0817 & 0.0048 \\
 & 5.15 & 5.22 & 5.30 & 0.1302 & 0.0041 \\
 & 5.02 & 5.08 & 5.15 & 0.0311 & 0.0046 \\
SDSS\,J1030+0524 & 5.98 & 6.07 & 6.16 & 0.0055 & 0.0013 \\[-2 pt]
 & 5.82 & 5.90 & 5.98 & 0.0140 & 0.0020 \\
 & 5.66 & 5.74 & 5.82 & 0.0114 & 0.0011 \\
 & 5.50 & 5.58 & 5.66 & 0.0444 & 0.0010 \\
 & 5.35 & 5.43 & 5.50 & 0.1219 & 0.0012 \\
 & 5.21 & 5.28 & 5.35 & 0.0561 & 0.0008 \\
SDSS\,J1137+3549 & 5.71 & 5.79 & 5.87 & 0.0056 & 0.0025 \\[-2 pt]
 & 5.56 & 5.63 & 5.71 & 0.0807 & 0.0017 \\
 & 5.40 & 5.48 & 5.56 & 0.1422 & 0.0020 \\
 & 5.26 & 5.33 & 5.40 & 0.0913 & 0.0017 \\
 & 5.12 & 5.19 & 5.26 & 0.1193 & 0.0017 \\
SDSS\,J1148+5251 & 5.77 & 5.86 & 5.94 & 0.0022 & 0.0003 \\[-2 pt]
 & 5.61 & 5.69 & 5.77 & 0.0109 & 0.0001 \\
 & 5.46 & 5.54 & 5.61 & 0.0123 & 0.0002 \\
 & 5.31 & 5.39 & 5.46 & 0.0422 & 0.0002 \\
SDSS\,J1250+3130 & 5.83 & 5.91 & 5.99 & -0.0212 & 0.0078 \\[-2 pt]
 & 5.67 & 5.75 & 5.83 & -0.0091 & 0.0054 \\
 & 5.51 & 5.59 & 5.67 & -0.0099 & 0.0034 \\
 & 5.36 & 5.44 & 5.51 & 0.0037 & 0.0042 \\
 & 5.22 & 5.29 & 5.36 & 0.0276 & 0.0030 \\
 & 5.08 & 5.15 & 5.22 & 0.0644 & 0.0027 \\
SDSS\,J1306+0356 & 5.72 & 5.80 & 5.89 & 0.0934 & 0.0013 \\[-2 pt]
 & 5.57 & 5.64 & 5.72 & 0.0495 & 0.0007 \\
 & 5.42 & 5.49 & 5.57 & 0.0491 & 0.0008 \\
 & 5.27 & 5.34 & 5.42 & 0.0436 & 0.0007 \\
 & 5.13 & 5.20 & 5.27 & 0.0630 & 0.0006 \\
 & 4.99 & 5.06 & 5.13 & 0.0785 & 0.0008 \\
ULAS\,J1319+0950 & 5.86 & 5.94 & 6.02 & 0.0019 & 0.0058 \\[-2 pt]
 & 5.70 & 5.78 & 5.86 & 0.0002 & 0.0049 \\
 & 5.54 & 5.62 & 5.70 & 0.0249 & 0.0029 \\
 & 5.39 & 5.46 & 5.54 & 0.0697 & 0.0036 \\
 & 5.24 & 5.31 & 5.39 & 0.0406 & 0.0027 \\
 & 5.10 & 5.17 & 5.24 & 0.1396 & 0.0029 \\
SDSS\,J1411+1217 & 5.63 & 5.71 & 5.79 & 0.0170 & 0.0008 \\[-2 pt]
 & 5.48 & 5.56 & 5.63 & 0.0355 & 0.0010 \\
 & 5.33 & 5.40 & 5.48 & 0.0543 & 0.0010 \\
 & 5.19 & 5.26 & 5.33 & 0.0384 & 0.0008 \\
 & 5.05 & 5.12 & 5.19 & 0.1211 & 0.0010 \\
 & 4.91 & 4.98 & 5.05 & 0.0429 & 0.0009 \\
SDSS\,J1602+4228 & 5.76 & 5.85 & 5.93 & 0.0242 & 0.0028 \\[-2 pt]
 & 5.61 & 5.68 & 5.76 & 0.0273 & 0.0013 \\
 & 5.45 & 5.53 & 5.61 & 0.0381 & 0.0014 \\
 & 5.30 & 5.38 & 5.45 & 0.0504 & 0.0012 \\
 & 5.16 & 5.23 & 5.30 & 0.0609 & 0.0009 \\
 & 5.02 & 5.09 & 5.16 & 0.0660 & 0.0011 \\
SDSS\,J1623+3112 & 5.95 & 6.04 & 6.12 & 0.0136 & 0.0031 \\[-2 pt]
 & 5.63 & 5.71 & 5.79 & 0.0276 & 0.0030 \\
 & 5.47 & 5.55 & 5.63 & 0.0383 & 0.0034 \\
 & 5.32 & 5.40 & 5.47 & 0.0198 & 0.0029 \\
 & 5.18 & 5.25 & 5.32 & 0.0225 & 0.0018 \\
SDSS\,J1630+4012 & 5.62 & 5.70 & 5.78 & -0.0062 & 0.0023 \\[-2 pt]
 & 5.47 & 5.54 & 5.62 & 0.0193 & 0.0024 \\
 & 5.32 & 5.39 & 5.47 & 0.1088 & 0.0028 \\
 & 5.17 & 5.24 & 5.32 & 0.0586 & 0.0020 \\
 & 5.04 & 5.10 & 5.17 & 0.1274 & 0.0028 \\
SDSS\,J2054-0005 & 5.78 & 5.86 & 5.95 & 0.0287 & 0.0039 \\[-2 pt]
 & 5.62 & 5.70 & 5.78 & 0.0150 & 0.0026 \\
 & 5.47 & 5.54 & 5.62 & 0.1131 & 0.0031 \\
 & 5.32 & 5.39 & 5.47 & 0.0652 & 0.0032 \\
 & 5.18 & 5.25 & 5.32 & 0.1196 & 0.0025 \\
 & 5.04 & 5.11 & 5.18 & 0.2436 & 0.0037 \\
SDSS\,J2315-0023 & 5.84 & 5.93 & 6.01 & -0.0127 & 0.0041 \\[-2 pt]
 & 5.53 & 5.60 & 5.68 & 0.0477 & 0.0037 \\
 & 5.38 & 5.45 & 5.53 & 0.0348 & 0.0042 \\
 & 5.23 & 5.30 & 5.38 & 0.1012 & 0.0031 \\
 & 5.09 & 5.16 & 5.23 & 0.1468 & 0.0031 \\
\enddata
\tablecomments{The different columns show the name of the object, the beginning of each redshift bin $z_{\rm start}$, the mean redshift of each bin $z_{\rm abs}$ and the end of $z_{\rm end}$ the redshift bin, and the mean flux of the continuum normalized spectrum with its uncertainty. 
\tablenotetext{a}{Note that the measurements along this quasar sightline have been replaced by the ones from \citet{Becker2015} in our master compilation.}}
\end{deluxetable}

\section{Spectral Bins Along All Quasar Sightlines}\label{sec:spec_bins}

In Fig.~\ref{fig:spec_bins_a}, \ref{fig:spec_bins_b}, and \ref{fig:spec_bins_c} we show all spectral bins of $50\,$\cmpch along all $23$ quasar sightlines in our data sample, indicating the respective optical depth measurements. 

\begin{figure*}[ht!]
\centering
\includegraphics[width=0.95\textwidth]{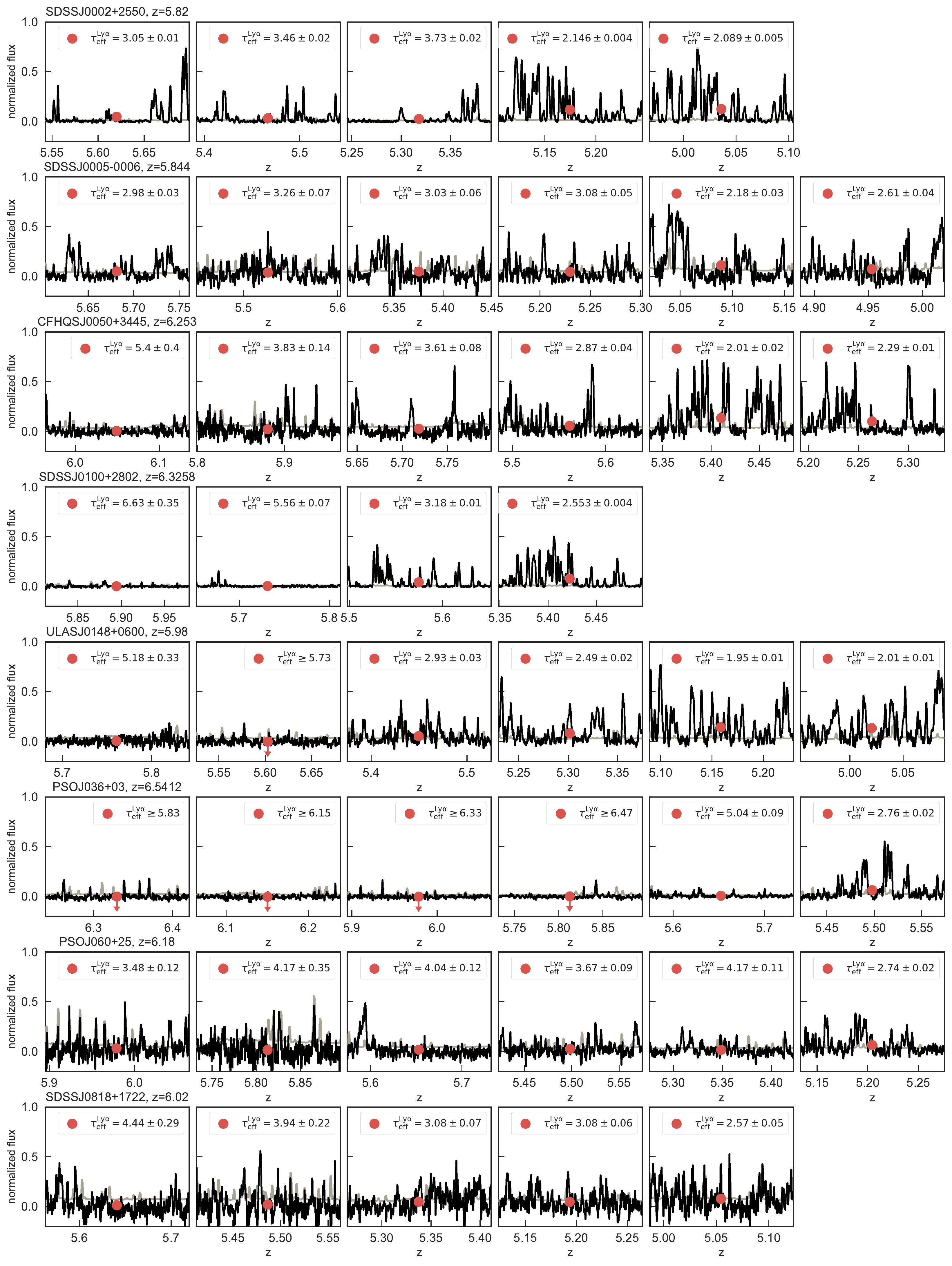}
\caption{Spectral bins of $50\,$\cmpch along all quasar sightlines in our data sample, for which we measure $\tau_{\rm eff}^{\rm Ly\alpha}$ within the \lya forest. The red data points show the measurements of $\langle F^{\rm obs}\rangle$ and the corresponding optical depth measurements are shown in the legend. \label{fig:spec_bins_a}} 
\end{figure*}

\begin{figure*}[ht!]
\centering
\includegraphics[width=0.95\textwidth]{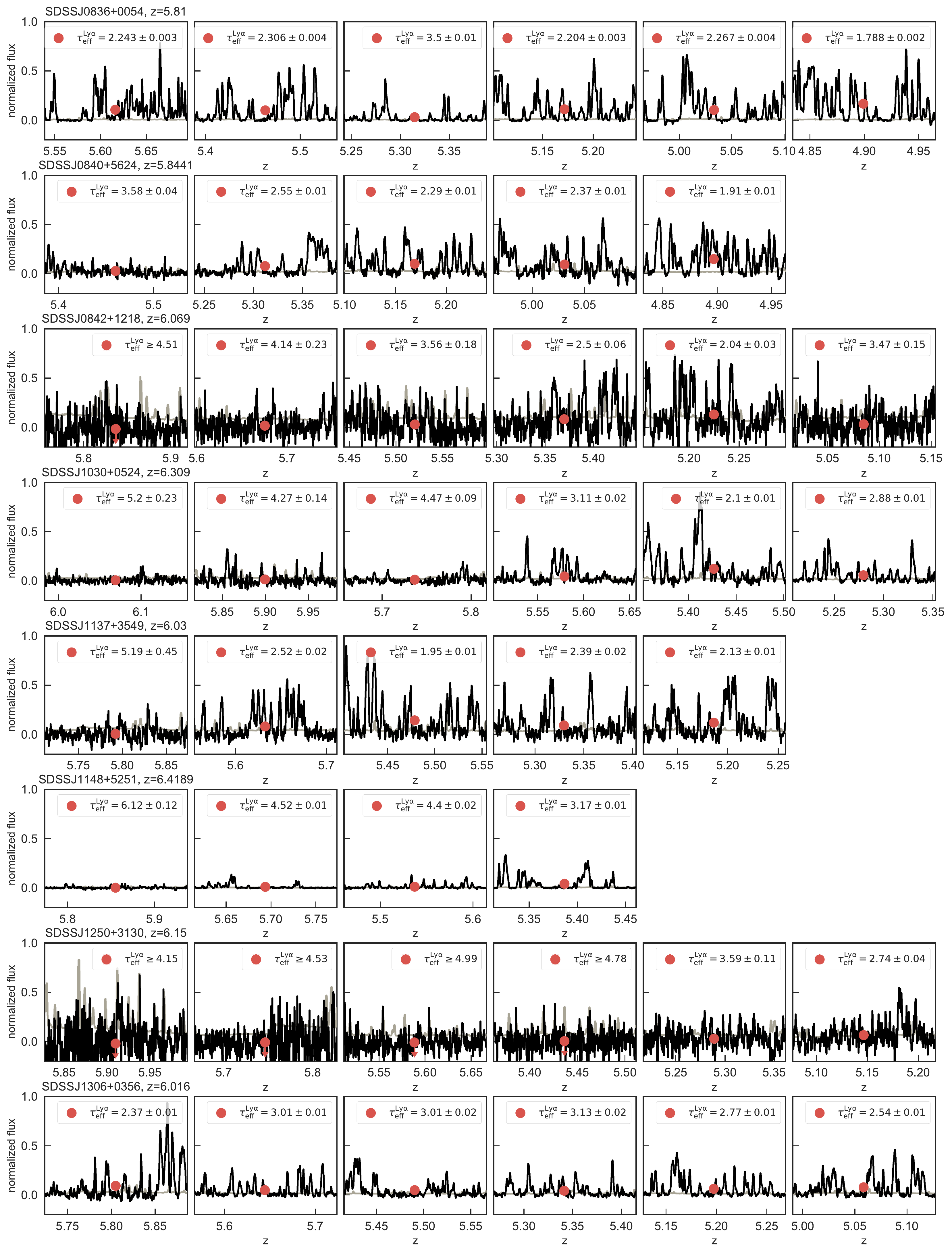}
\caption{Continuation of Fig.~\ref{fig:spec_bins_a}. \label{fig:spec_bins_b}} 
\end{figure*}

\begin{figure*}[ht!]
\centering
\includegraphics[width=0.95\textwidth]{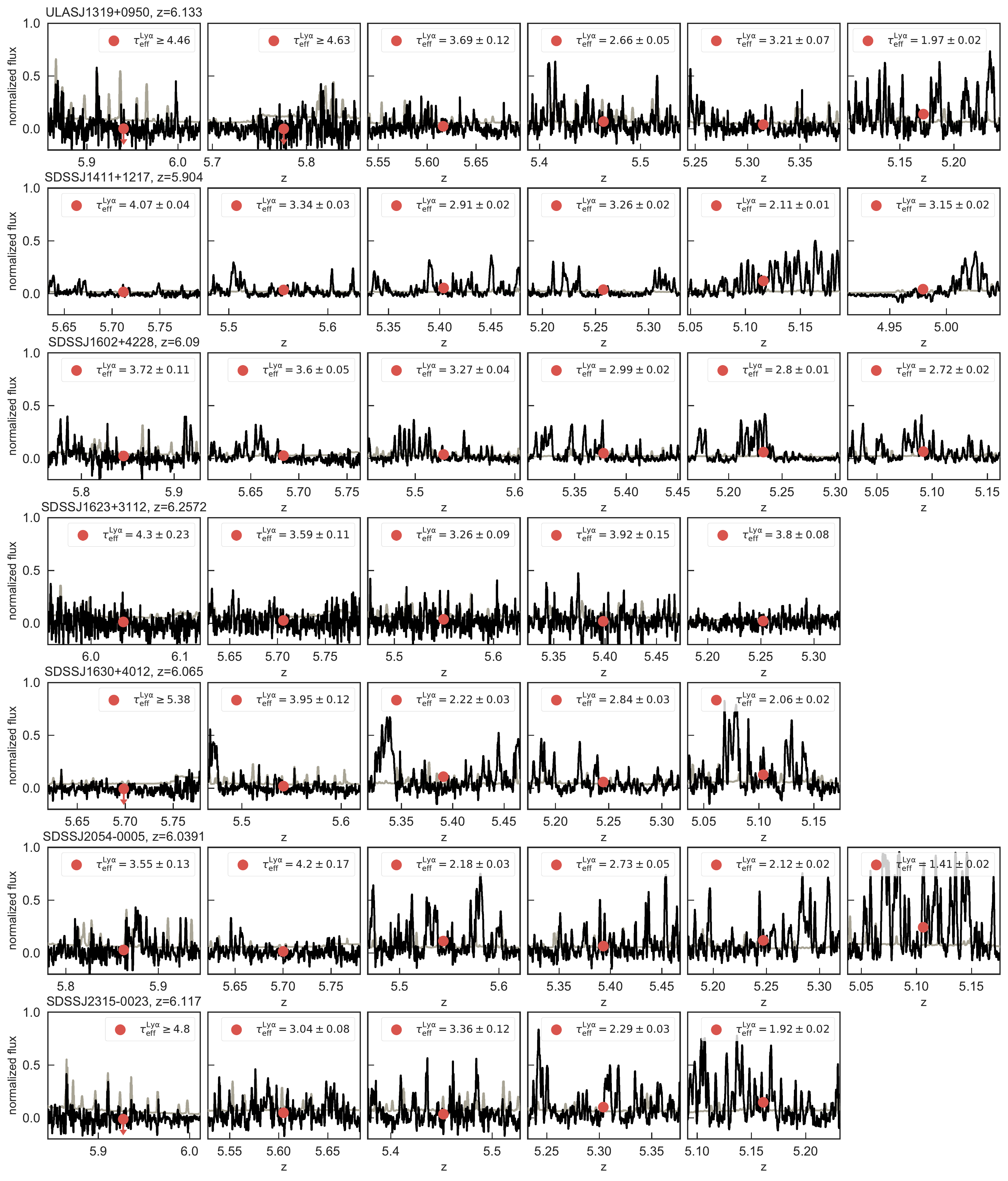}
\caption{Continuation of Fig.~\ref{fig:spec_bins_a} and Fig.~\ref{fig:spec_bins_b}. \label{fig:spec_bins_c}} 
\end{figure*}

\end{document}